\documentstyle[preprint,aps]{revtex}

\newcommand{\Sp}{\mbox{Tr}}
\newcommand{\Tr}{\mbox{Tr}}
\newcommand{\1}{\openone}

\renewcommand{\+}{\uparrow }
\renewcommand{\-}{\downarrow }

\begin{document}

\title{\large \bf  FUNCTIONAL INTEGRALS FOR HUBBARD OPERATORS\\
AND \\
PROJECTION METHODS FOR
STRONG  INTERACTION}
\author{Eberhard O. T\"ungler and Thilo Kopp }
\address{Institut f\"ur Theorie der Kondensierten
Materie, Universit\"at Karlsruhe,
76128 Karlsruhe, Germany}
\date{\today}
\maketitle
\begin{abstract}
We discuss  problems  of  functional integral
formalisms in a constrained fermionic Fock space.
A functional integral is set up for the Hubbard model
using generalized coherent states which lie either in the constrained
or in the full Fock space.
The projection for the latter states is implemented
through a reduction of the charge fluctuations which induce transitions
between the constrained and full space.
The Lagrangian is expressed in terms of two complex fields representing spin
and
charge excitations,
and one Grassmann field
signifying   hole excitations. Here, the charge excitations denote transitions
between states with empty and doubly occupied sites.
The projection method is inspired by the observation that the local interaction
in the model resembles a magnetic field in the space of charge
fluctuations. Hence the projection is understood as an infinite magnetic
field in a  spin path integral.
\end{abstract}

\pacs{PACS numbers here}

\section{Introduction}

 Models of interacting fermions on the lattice like the Hubbard model
and its derivatives,
the $t-J$ and Heisenberg models, have been under intense investigation
in recent years \cite{Fradkin}.
These models are thought to describe much of the qualitative physics
of the Mott-Hubbard
metal-insulator transition, including the magnetic
correlations and possibly superconductivity in its vicinity.
Most importantly, there is reason to expect new types of ground states
such as spin liquid states \cite{Anderson} in the two-dimensional systems.
Progress in assessing the validity and relevance to many  existing
ideas proposed in this context has been  slow due to the lack of reliable
theoretical methods.
The generic problem here is the dynamics of fermions in constrained
Fock  space,
excluding double occupancy of lattice sites.
A convenient representation used in this case is in terms
of slave particles \cite{Barnes},
assigning fermion and boson creation operators to the various
possible
states at a given lattice site. The occupation numbers of the slave
particles at any given site     then have to add up to one for all times.
Local constraints of this type are difficult to handle.
Most treatments use a mean field description by which the
constraint is only satisfied on the average\cite{Read2,Read3}. Additionally,
the slave boson methods exhibit  a local gauge invariance which is
destroyed in the mean field description.
The conserving slave boson approach presented in  \cite{Woelfle},
however, preserves the local gauge invariance. \\

One strategy to avoid slave particles
builds on a Hubbard Stratonovich decoupling of the
interaction in terms of density and spin excitations. But a straightforward
saddle point evaluation
violates certain symmetries, e.g.\   the spin rotation invariance.
Schulz \cite{Schulz} considered a time dependent quantization axis for
each spin which is being treated as an
additional set of angular field variables.
This procedure preserves the spin rotation invariance of
the saddle point Lagrangian.

A different strategy is to try to work in the
constrained subspace only.
Conventional many body formulations  build on equation of motion and
diagrammatic schemes \cite{Hubbard,Zaitsev,Rucki},
and often involve uncontrolled
factorizations which  can    violate sum rules and
conservation laws (except for
certain limiting
cases, as the   limit of large spin degeneracy in \cite{Rucki}).
The use of resolvent operator methods \cite{Fulde} may be
well adapted to certain problems.
Also Quantum Monte-Carlo techniques, which work in the constrained subspace
only,
were devised in   \cite{Hewson,xyzhang}.

In this paper we explore  the following alternative:
the construction of a path integral formalism  for   Hubbard
operators \cite{Wiegmann,Weller}, different
from the usual fermionic representation \cite{sss}.
The path integral formalism has several well-known properties: \\
(i) it allows   a time-ordered perturbation theory to be developed\\
(ii) collective behavior may be incorporated from the outset \\
(iii) the classical limit may be easily extracted, e.g.\ in
the case of large spin degeneracy. \\
In addition the path integral we develop here does not exhibit the   local
gauge
invariance generated in slave boson approaches.

We organize the paper as follows:\\
First,  we  introduce   functional integral
formulations in the constrained Fock space.
We note some problems in the literature concerning the measure.
They may be traced back to  difficulties with the
resolution of the unity with normalized coherent states.
Therefore we review the definition of coherent states in order to
extend it to graded algebras.
We derive  coherent states in the full Fock space using
two complex fields in order to describe the spin and  charge
degrees of freedom, and one Grassmann field  in order
to  implement
fermionic statistics. Working with these coherent states, we are
able to   resolve
the unity  and  set up
a functional integral with  the Trotter formula
for which the measure is known exactly.
In order to test   its validity we  calculate the so-called atomic limit.
To handle the problem of a constrained space we develop a
well-defined projection method, first  for spin $1/2$
functional integrals\cite{Klauder,Stone}
in strong magnetic fields,  and later   adapting it to  the   electron
path integral for strong local interaction.
We  investigate the Hubbard model in  the
limit of strong  Coulomb interaction to derive an effective Lagrangian.

\section{Functional Integral Formulations Of The Hubbard Model }

\noindent As a model system we consider a fermionic lattice Hamiltonian
with local interactions which is most suitable for a discussion of
projection methods.  The Hubbard Hamiltonian \cite{Hubbard,Gutzwiller}
allows   ``to turn on'' the projection  by sending the interaction parameter
to infinity.
Using canonical electron creation and annihilation operators it reads:
\begin{eqnarray}
H=-\sum _{\sigma =\left\{ \+,\- \right\} }
\sum _{\langle i,j \rangle } \  t_{ij} \ C^{\dagger} _{i,\sigma }C _{j,\sigma }
 +U \sum _{i} \  C^{\dagger} _{i,\+}C _{i,\+}C^{\dagger} _{i,\-}C _{i,\-}
\nonumber
-\mu \sum _{\sigma ,i}C^{\dagger}_{i,\sigma }C_{i,\sigma }
\end{eqnarray}
where $\langle i,j \rangle $ is a summation over nearest neighbor sites,
and we will assume that $t_{ij} = t $ for all nearest neighbors
on a hypercubic lattice.

This model has a graded structure in the following sense:\\
local excitations of the system either belong to a bosonic or fermionic type.
As
bosonic excitations we classify all excitations involving
an even number of electron creation and annihilation operators.
Spin, particle hole and pair excitations are in this class and will
be attributed to
complex fields in the path integral.
Fermionic excitations, with an odd number  of electron
operators,  are creation or annihilation of electrons at empty, singly
or doubly occupied sites, and will be taken care of by Grassmann fields.

To set up this formalism on the Hamiltonian level first, we
introduce local operators which represent these excitations
more faithfully.
Knowing the algebra of  these operators (the ``Hubbard operators''), one is
in the position to derive coherent states for the  excitations.
These Hubbard operators are defined as projections:
\begin {eqnarray}
X^{\mu \nu }_{i} := | i;\mu \rangle \langle i;\nu |
\end {eqnarray}
where $| i;\mu \rangle $ are orthonormal states at site $i$,
representing the empty  ($\mu=0$),    singly  occupied
($\mu = \+,\-$) and doubly ( $\mu =2 $ ) occupied sites (we chose the
convention
$|2\rangle =C^{\dagger } _{\-} C^{\dagger } _{\+}|0\rangle$). Then one can
verify
the following supercommutator relation:
\begin {eqnarray}
\left[ X^{\mu \nu }_{i} , X^{\alpha \beta }_{j}\right]_{S}=
\delta _{ij}\left(
X^{\mu \beta  }_{i}\delta _{\nu \alpha}
- \left( -1 \right)
^{\acute \chi ^{\mu \nu }_{i}\acute \chi ^{\alpha \beta }_{j}}
X^{ \alpha \nu  }_{i}
\delta _{\beta \mu } \right)   \label{Superkommutator}
\end {eqnarray}
where $S=2\left( \Theta (\acute \chi ^{\mu \nu } +
\acute \chi ^{\alpha  \beta  } -3/2) - 1/2 \right) $,
and the graded characters $\acute \chi ^{\mu \nu }$ of the
Hubbard  operators are 0 in the case of the bosonic  operators
($ X^{00}, X^{2 2}, X^{\sigma \sigma }, X^{\sigma -\sigma },
X^{02}$, $ X^{2 0} $) and 1 in the case of the fermionic operators
$ X^{0\sigma }, X^{\sigma 0}, X^{2\sigma }$, $ X^{\sigma 2 }$.
The algebra of the Hubbard operators has been extensively
studied in \cite{Zaitsev}.

Now we may write the canonical creation and annihilation  operators
in terms of linear combinations of Hubbard operators:
\begin {eqnarray}
C_{\sigma }& =& X^{0 \sigma }+\sigma X^{-\sigma 2 }\nonumber \\
C^{\dagger} _{\sigma } & =& X^{ \sigma 0 }+\sigma X^{2 -\sigma  }
\end {eqnarray}
However, this transformation is not linear and the inverse of it reads:
\begin {eqnarray}
\begin {array}{ll}
X^{\sigma 0} \ \,   = C^{\dagger} _{\sigma } \left(
1-n_{-\sigma}\right)  &
X^{0 \sigma }  \ \,   = C_{\sigma } \left( 1-n_{-\sigma}\right)  \\
X^{\sigma 2}  \  \,  = C_{- \sigma }   n_{\sigma}   &
X^{2 \sigma }  \  \,  = C^{\dagger} _{-\sigma }   n_{\sigma}  \\
X^{\sigma \sigma }  \  \,  =  n_{\sigma }\left(1- n_{-\sigma}\right)  &
X^{\sigma -\sigma }    = C^{\dagger}_{\sigma } C_{-\sigma }   \\
X^{0 0}  \ \,   = \left(1- n_{\sigma}\right)\left(1- n_{-\sigma}\right)    &
X^{2 2} \  \,   =  n_{\sigma} n_{-\sigma}  \\
X^{0 2}  \ \,   =   C_{ \+ } C_{ \- }  &
X^{2 0} \  \,   =   C_{ \- }^{\dagger} C_{ \+ }  ^{\dagger}
\end {array}
\end {eqnarray}
The bosonic number operators automatically satisfy the
local completeness relation $ \1 = X^{00} +  X^{\+ \+  } +X^{\-\-}+X^{22}$
which follows from the orthogonality and completeness of the local states.

Furthermore, the Hubbard Hamiltonian takes the form:
\begin{eqnarray}
H=- t \sum _{<ij>} \sum _{\sigma }
(X^{ \sigma 0 }_{i } +\sigma X^{2 -\sigma  }_{i })
(X^{ 0 \sigma }_{j } +\sigma X^{ -\sigma  2}_{j })
-\mu \sum _{i }\sum _{\sigma } X^{ \sigma \sigma  }_{i }
+(U-2\mu ) \sum _{i} X^{22}_{i}  \label{ham}
\end{eqnarray}
Obviously, the kinetic energy is split into four terms, each of them
representing
a specific interaction process between excitations on nearest neighbor sites.
The local terms in this Hamiltonian may be interpreted as chemical potentials
$\mu$
and $2\mu -U$ for singly and doubly  occupied sites, respectively.

Now it seems very easy to project out
doubly occupied states with these Hubbard operators .
Since the last term in (\ref{ham}) attributes an exponentially small weight to
doubly occupied sites for large
$U$ it  appears to be obvious  that the infinite $U$ limit is accomplished
by discarding all terms with
index 2 in equation (\ref{ham}).
We will show that in the Lagrangian formalism, i.e.\ in
the path integral formalism, such a projection scheme fails -- except
for very exotic choices of the measure of the path integration.  \\

Construction of a path integral for the degrees of freedom
represented by the Hubbard operators requires a mapping onto classical
fields or Grassmann fields depending on the statistics.
Wiegmann \cite{Wiegmann} formulated the path integral representation
using
geometric quantization.   First, he introduced the projection operator
$Q$ onto coherent
states.
The  Lagrangian is:
\begin{eqnarray}
L=-\Sp (HQ) + \frac{1}{2}\int _{0} ^{1} d u \ \Sp (Q \ \partial _{u} Q \
\partial
_{\tau } Q) \label{Lagrangian1}
\end{eqnarray}
where $Q(\tau , u)$  is an arbitrary differentiable extension
of $Q(\tau )$---only restricted by  the
boundary conditions
$Q(\tau , 0 )=const $ and $ Q(\tau , 1 )=Q(\tau )$ and
$\tau , u$ are  imaginary time variables.
The second term in (\ref{Lagrangian1}) is
the so-called Berry phase.
This  phase is being collected by the system while transporting it
along a considered  path \cite{Berry}.
The first part, containing the Hamiltonian of (\ref{Lagrangian1}),
is the `Boltzmann weight' of this path.

In order to
investigate the limit of infinitely strong
local interaction $(U=\infty)$ Wiegmann proposed the following form for
$Q(\tau)$:
\begin{eqnarray}
Q(\tau)= (1-\psi ^{\dagger} \psi  )\sum _{\sigma \sigma '}
A_{\sigma \sigma '}X^{\sigma \sigma '}
+\sum _{\sigma }\psi ^{\dagger}z_{\sigma } X^{\sigma 0} +
\sum _{\sigma } X^{0 \sigma } z_{\sigma }^{\dagger} \psi
+ \psi ^{\dagger} \psi X^{00} \label{lll1}
\end{eqnarray}
where
\begin{eqnarray}
 z_{\sigma } =\left(
\begin{array}{l}
\sin (\vartheta )e^{i \varphi }\\
\cos  (\vartheta )
\end{array}
\right)
\ \ \mbox{and} \ \
A_{\sigma \sigma '}=
\left(
\begin{array}{ll}
\sin ^{2} (\vartheta ) &  \sin  (\vartheta ) \cos (\vartheta)e^{i \varphi }\\
\sin  (\vartheta ) \cos (\vartheta )e^{-i \varphi } &
\cos  ^{2} (\vartheta )
\end{array}
\right) \nonumber
\end{eqnarray}
and $\psi ^{\dagger }, \psi$ are Grassmann numbers.
Only empty and singly occupied  states are involved
in (\ref{lll1}), making it plausible that a projection
onto a constrained space has been successfully implemented.
Using (\ref{Lagrangian1}) and (\ref{lll1}), the partition function is:
\begin{eqnarray}
Z = \Tr e^{-\beta H} &=& \int {\cal D } \left[  \vartheta ( \tau ),
\varphi (\tau ), \psi ^{\dagger} (\tau  ),  \psi  (\tau  ) \right]   \\
& & \times \exp \left[ -\int _{0}^{\beta }\left(
-i\dot{\varphi }\sin^{2}{\vartheta } +
\psi^{\dagger} \left( \partial _{\tau }
+i \dot{\varphi}\sin^{2} {\vartheta }
\right)
\psi
+\Sp (QH)
\right) d \tau \right] \nonumber
\end{eqnarray}
To point out the  problem with the normalizability of $Z$
we consider the following  trivial limit $\Sp (Q H )=0$.
In this case the partition function should be equal to the dimension
of the Fock space.
Applying the relations of appendix \ref{Anhang2}, it is straightforward
to integrate out the
Grassmann degrees of freedom,   and we find:
\begin{eqnarray}
Z =  \int {\cal D } \left[  \vartheta ( \tau ),
\varphi (\tau ) \right] \ \
\left[ 1+
\exp \left[ \int _{0}^{\beta }d \tau  \ i\dot{\varphi
}\sin^{2}{\vartheta } \right]
\right]=\lim _{N \rightarrow \infty } (2^{N} + 2 )
\end{eqnarray}
using the results of
appendix  \ref{Anh1}, where $N$ is the number of time
slices in the discretized partition function.  This partition function
cannot be normalized to the correct result, $1+2$, with  a common
prefactor in the spirit of Feynman's path integral.

To further illuminate this problem, we  investigate the measure.
It should be possible to resolve  the unity operator with coherent states:
\begin{eqnarray}
\int _{0}^{\pi }  d \vartheta  \int _{0}^{2 \pi }  d \varphi
\int d \psi  \int d \psi ^{\dagger} \ \mu (\vartheta ,\varphi ,\psi , \psi
^{\dagger} )\
Q= X^{00} \, + X^{\+ \+ } \, + X^{\- \- } =\1, \label{zzzzzz}
\end{eqnarray}
where $\mu $ is a measure   to be determined. It can only be a function of
the spin variables, otherwise the
term involving $X^{00}$ would be eliminated.
 From  (\ref{lll1}) the
prefactor of $X^{00}$ in (\ref{zzzzzz}) is:
\begin{eqnarray}
\int _{0}^{\pi }  d \vartheta  \int _{0}^{2 \pi }  d \varphi
\ \mu (\vartheta, \varphi ) =1  \label{ffg}
\end{eqnarray}
and accordingly of  $X^{\+ \+}, X^{\-\-}$:
\begin{eqnarray}
\int _{0}^{\pi }  d \vartheta  \int _{0}^{2 \pi }  d \varphi
\ \mu (\vartheta, \varphi )\sin ^{2} (\vartheta /2) =1 \ \ \ \
\int _{0}^{\pi }  d \vartheta  \int _{0}^{2 \pi }  d \varphi
\ \mu (\vartheta, \varphi )\cos ^{2} (\vartheta /2) =1,
\label{Widerspruch}
\end{eqnarray}

Adding the two last equations results in  a contradiction to (\ref{ffg}).
As a consequence, the resolution of the unity operator is not possible with $Q$
of equation (\ref{lll1}) \footnote{It seems, the doubly occupied sites
have not been projected out consistently. }. However, the
derivation of (\ref{Lagrangian1}) requires such a resolution. \\

Weller \cite{Weller}  was  aware of this  problem and succeeded
in setting up   a resolution of the unity operator.
He obtained   the following relation:
\begin{eqnarray}
\int \ Q &=&\int \ \ \left[
1+\frac{1}{4} \psi \psi ^{\dagger} +\psi \sin (\vartheta /2 ) e^{i\varphi }
X^{\+ 0} + \psi \cos ( \vartheta /2) X^{\- 0}
\right] | 0\rangle \nonumber \\
& & \ \cdot \langle 0|
\left[
1+\frac{1}{4} \psi \psi ^{\dagger} +X^{0\+ }\psi ^{\dagger} \sin (\vartheta /2
)
e^{-i\varphi }
 + X^{0\- }\psi \cos ( \vartheta /2 )
\right] \nonumber \\
& & = X^{00} \ +\ X^{\+ \+ } \ +\ X^{\- \- } \ =\1 \label{lala3}
\end{eqnarray}
where
\begin{eqnarray}
\int \ = \ 2\int _{0}^{\pi} \frac{d \vartheta }{\pi }
\int _{0}^{2\pi} \frac{d \varphi  }{2\pi } \int d\psi ^{\dagger}
\int d\psi
\end{eqnarray}
But $Q$ of \cite{Weller} is not a projector, i.\ e.\ $Q^{2} \ne Q$.
In other words the coherent state is not normalized. Using
the relation (\ref{lala3}),  Weller  derived a path integral.
The  resulting  Berry phase has an unusual form:
\begin{eqnarray}
S_{0}&=&\frac{1}{2}  \sum _{n=1}^{N} \psi ^{\dagger} _{n} \psi _{n} -
\sum _{n=1}^{N} \psi^{\dagger} _{n}\psi _{n-1}\times \nonumber \\& &\times
\left(
\sin(\vartheta _{n}/2) \sin(\vartheta _{n-1}/2) e^{i(\varphi _{n}
- \varphi  _{n-1})}
+\cos(\vartheta  _{n}/2) \cos(\vartheta _{n-1}/2)
\right) \nonumber \\
&\rightarrow & \int _{0} ^{\beta} d \tau \
\psi^{\dagger} \left( \partial _{\tau }
-i \dot{\varphi}\sin^{2} (\vartheta /2)
\right)
\psi
-\frac{1}{2} \sum _{n=1}^{N}  \psi _{n} ^{\dagger}  \psi _{n}
\end{eqnarray}
It is impossible to take  the continuum limit because the last term
is missing a  prefactor
$\epsilon =\beta / N$.
The Hamiltonian part, $Tr (HQ)$,  has the same form as
in Wiegmann's paper.
One can prove
that a resolution of the unity operator with normalized coherent states is not
possible in the case of the $U=\infty $ algebra (see appendix \ref{klkl}). \\

To investigate the $U=\infty $ limit,  Schmeltzer \cite{Schmeltzer}
used a slave boson
technique
in the full Fock space:
\begin{eqnarray}
L=L_{0}+L_{\lambda}+L_{H}
\end{eqnarray}
with
\begin{eqnarray}
L_{0}&=&\sum _{i} \left[
\psi ^{\dagger}_{i}  \partial _{\tau } \psi _{i}  +
\sum _{\sigma }b^{\dagger}_{i,\sigma }\partial _{\tau }b_{i,\sigma }
\right]
\nonumber \\
L_{\lambda} &=& \sum _{i} i \tilde{\lambda } _{i}
\left(  \psi ^{\dagger}_{i}\psi _{i} + \sum _{\sigma }
b^{\dagger}_{i,\sigma }b_{i,\sigma } -1\right) \nonumber \\
L_{H}  &=&
\sum _{i}\left[ \mu \psi ^{\dagger}_{i} \psi _{i} - \mu \right]
-\sum _{\sigma } \sum _{<i,j>}
\left( t_{ij} \psi ^{\dagger}_{i} b^{\dagger}_{i\sigma }b_{j\sigma }\psi _{j}
+ \mbox{H.c.} \right)
\end{eqnarray}
where   $b_{\sigma
} $ are
complex fields representing the spin degrees of freedom,   $\tilde{\lambda } $
is a Lagrange multiplier field,
and $\psi $ is a Grassmann field  for the hole excitation.

A graded transformation  was set up
\begin{eqnarray}
b_{\sigma}=z_{\sigma }\sqrt{1- \psi ^{\dagger}\psi   } \ \ \ \
\lambda=\tilde{\lambda } \left( 1- \psi ^{\dagger}\psi  \right)
\end{eqnarray}
which transforms the graded constraint  into a trivial complex
constraint
which can be solved by parameterization.
Consequently, the Lagrangian is $L=L_{0}+L_{\lambda}+L_{H}$, where
\begin{eqnarray}
L_{0}&=&\sum _{i} \left[ \psi ^{\dagger} _{i}\partial _{\tau } \psi _{i}+
\left( z^{\dagger}_{i,\sigma }\partial _{\tau }z_{i,\sigma }
\right)
\left( 1- \psi ^{\dagger}_{i}\psi  _{i}\right) \right] \nonumber \\
L_{\lambda}&=& \sum _{i}i\lambda _{i}\left( \sum _{\sigma }
z^{\dagger}_{i,\sigma }z_{i,\sigma }
-1\right) \nonumber \\
L_{H}&=&  \sum _{i} \left[ \mu \psi ^{\dagger}_{i}\psi_{i}-\mu \right]
+\sum _{<i,j>} \left( t_{ij}\psi ^{\dagger}_{i} z^{\dagger}_{i\sigma }
z_{j\sigma }\psi _{j} +\mbox{H.c.} \right)
\end{eqnarray}
The resulting
Lagrangian is identical to  Wiegmann's, and therefore suffers from
the same problem of normalizability.  \\

Schulz \cite{Schulz} also worked out a path integral formalism for the  full
Fock space. He started with the  canonical  formalism and decoupled the
model  by a Hubbard Stratonovich transformation.
Having introduced a unity at each space and time step, he chose
the unity to be $\1= RR^{\dagger}$, where $R$ is a spin rotation matrix
$SU(2)/U(1)$. In order to make $R$ a   dynamical field, he additionally
integrated
over all angle
variables at each space and time step using the invariant measure of
the group.
In this way  he built in a dynamical, rotating spin reference system.
After taking a  saddle point approximation with second order
fluctuations, he  applied a $1/U $ expansion and integrated out the
Hubbard Stratonovich field variables.
Again the effective Lagrangian  determined in this  way is equivalent to
Wiegmann's in  the limit $U=\infty $.
Also Weng et al.\cite{Ting}    introduced  a
dynamical rotating spin reference system in the same manner as Schulz.

\section{Coherent States}
\label{new}

In the following sections we will   derive a path integral formalism
which uses  only one Grassmann field variable and  works   with
a well-defined
measure.
It should give a correct result for the partition function and, in addition,
the measure of the path integral should  possess a  controlled continuum
limit in the time direction.

First,  we   reconsider  how to set up the definition of coherent states.
A well  known procedure  is  to require the coherent state to be an eigenstate
of an annihilation operator.
This  results in an unnormalized state (see \cite{Negele}).
A straightforward generalization of this scheme to an arbitrary
operator algebra is unclear.
Therefore we take the definition of Perelomov \cite{Perelomov,Zhang}
for generalized coherent states.
The underlying idea is that the coherent states  are  the orbit of a Lie group
represented in a Hilbert space and acting on a reference  state.
This scheme produces normalized coherent states. We will discuss
only those coherent
states which fulfill the two
conditions:
(I) normalization and (II) resolution of the unity operator.
This will automatically result in a well-defined Berry phase in the
continuum limit\footnote{Abandoning condition (I) may
result in a Berry phase which has no continuum limit, cf.\
Weller \cite{Weller}.}.

The limit of infinite local interaction is of special interest in this paper.
It should be  advantageous to construct normalized coherent states
in the constrained space.
However, we will show in appendix \ref{klkl}
that possible  coherent states with one Grassmann field  in the constrained
space will not fulfill
(I) and (II) simultaneously.

For the coherent state to be defined  uniquely,  up to a phase factor, the
Lie group should be divided in left cosets of the   maximal isotropic subgroup,
i.e.\
all operators, which when applied to the reference state generate just a phase
factor, are removed.

In the   case of bosons the coherent state is:
\begin{eqnarray}
e^{\alpha  a^{\dagger} - a \alpha ^{\dagger}}|0\rangle \label{gl1}
\end{eqnarray}
and in the fermionic case:
\begin{eqnarray}
e^{\psi  C^{\dagger} - C \psi ^{\dagger}}|0\rangle \label{gl2}
\end{eqnarray}
where $\alpha $ is a complex variable, $\psi $ a Grassmann number,
$a,a^{\dagger}$ bosonic creation and annihilation operators and
$C,C^{\dagger}$ the fermionic operators, respectively.
The unity operator $\1$  does not appear in  the exponentials of  equations
(\ref{gl1})
and (\ref{gl2}) because it is a generator
of the maxumal isotropic subgroup.

The graded  case  is more involved.
As mentioned above,  bosonic and fermionic Hubbard operators
are the generators of the graded Lie group.
To get the new coherent state $| G \rangle  $ we write the analogous ansatz:
\begin {eqnarray}
| G \rangle =\exp \left[ { \sum _{a\neq b}  \chi _{ab} X^{ab}
- h.c. }\right]
| \+ \rangle \label{DefkohZust}
\end {eqnarray}
where we have just chosen $| \+ \rangle $ to be the arbitrary reference
state\footnote{The discussion of Klauder \cite{Klauder} concerning the correct
choice
of the reference state is not applicable here because both
$ | \+  \rangle $ and
$|\downarrow \rangle $ are states with maximal weight.}.
Now, we intend  to use only one Grassmann field and the smallest number
of complex fields   possible.
The minimal number of complex fields is two.
This expresses the fact that we deal with a spin and a charge degree of
freedom.
Therefore, we choose
$\chi _{\+ 0}=\alpha \psi ,\ \ \chi _{0\+ }=\alpha ^{\dagger} \psi ^{\dagger},
\ \ \chi
_{\-0}=\beta  \psi
, \ \ \chi _{0\-}=\beta  ^{\dagger} \psi ^{\dagger}$
and $\alpha , \beta , \chi _{ab}$ to be complex. Then  the
exponential function  is expanded using $\psi ^{2}=0,$ etc.\ :
\begin{eqnarray}
|G \rangle &=& \left(  1 + \sum _{ab } \chi _{ab} X_{ab}
+ \frac{1}{2}\sum _{a\neq b }\sum _{c\neq d }\chi _{ab}\chi _{cd} X_{ab}
X_{cd}
+ ...
\right) | \+ \rangle \nonumber  \\
&=& \left( a+ b \psi \psi ^{\dagger} \right) | \+ \rangle
+\left( c+ d \psi \psi ^{\dagger} \right) | \- \rangle
+ e \psi \  | 2  \rangle
+ f \psi \ | 0 \rangle   \label{kohZustansatz}
\end{eqnarray}
where $a,b,c,d,e,f$ are complex variables which are   functions of the
$\chi _{ab}$\footnote{In appendix   \ref{Widerspruchsbeweis} we will
show the calculation
of these functions explicitly in two special cases.}.
The conditions to be fulfilled by  the coherent state are $(I)$
normalization and $(II)$ existence of a resolution of the unity operator:
\begin {eqnarray}
\begin {array}{lrcl}
(I)      & \langle G | G \rangle &= &
 a^{\dagger} a +a^{\dagger} b \psi \psi ^{\dagger} +a b ^{\dagger}\psi \psi
^{\dagger}
+c^{\dagger} c  +c d ^{\dagger} \psi \psi ^{\dagger} +
c^{\dagger} d \psi \psi ^{\dagger} +
e^{\dagger} e \psi ^{\dagger} \psi + f^{\dagger}f \psi ^{\dagger} \psi
= 1
\label{a}\\ & & & \\
(II)    &  \1& =& \int d \mu \left( a,b,c,d,e,f \right) d \psi ^{\dagger} d
\psi
|G \rangle
\langle G|   \\
&   & = &
\int d \mu \left( a,b,c,d,e,f \right)
 \left( \left(b a^{\dagger}+ab^{\dagger}\right) \right. X^{\+ \+}
+|e|^{2}X^{22 }
+|f|^{2}X^{00} +
\left( cd^{\dagger} +   dc^{\dagger} \right) X^{\- \-}  \\
&&\stackrel{!}{=} &
X^{00}+X^{\+ \+ }+X^{\-\-}+X^{22} = \1 \label{b}
\end {array}\nonumber
\end {eqnarray}
\footnote{ The integration over the phases of $cd^{\dagger}$ and
$(af^{\dagger}+be^{\dagger})$
which are the prefactors  of the offdiagonal
Hubbard operators are supposed to give 0. }
Here, $\mu $ is an
unknown measure which still has to be determined:
The prefactors of the diagonal Hubbard operators are integrals and
should   equal  1:
\begin{eqnarray}
\begin{array}{ll}
\int d \mu \left( a,b,c,d,e,f \right) \left( b a^{\dagger}+ab^{\dagger} \right)
\stackrel{!}{=} 1
& \int d \mu \left( a,b,c,d,e,f \right)  |e|^{2} \stackrel{!}{=} 1 \\
\int d \mu \left( a,b,c,d,e,f \right)  |f|^{2} \stackrel{!}{=} 1 &
\int d \mu \left( a,b,c,d,e,f \right)    \left(
cd^{\dagger}+dc^{\dagger}\right) \stackrel{!}{=} 1
\end{array} \label{jkjkjk}
\end{eqnarray}
In addition, the normalization
condition $(I)$ has to be fulfilled.
The solution of this  system of equations is not unique  since  there are
six unknown variables and one unknown measure function $\mu $.
One solution is, e.g.\ :
\begin {eqnarray}
\begin {array}{rl}
| G \rangle =
\left( 1+ \frac{1}{2} \psi \psi ^{\dagger} \right) \sin  \left( \vartheta
\right)e^{i \varphi }& | \+  \rangle  \label{kohZust}\\
+\left( 1+ \frac{1}{2} \psi \psi ^{\dagger} \right) \cos \left( \vartheta
\right) \ \ &| \-  \rangle    \\
+ \  \psi  \cos \left( \theta \right) \ \ \ &| 0 \rangle   \\
+ \   \psi  \sin  \left( \theta  \right)e^{i \phi } &| 2 \rangle
\end {array} \label{ioioio}
\end {eqnarray}
This state is normalized and one can resolve the unity operator and
obtain a trace:
\begin {eqnarray}
\1 =\int  \
|G \rangle \langle G |, \hspace{2cm}
\Sp \left( A \right)= \int  \
\langle {\scriptstyle \chi } G  |A| G \rangle
\label{Aufloesungdes}
\end {eqnarray}
where $\int =  2
\int _{0} ^{\pi} \frac{d \theta }{\pi }
\int _{0} ^{2\pi} \frac{d \phi }{2\pi }
\int _{0} ^{\pi} \frac{d \vartheta }{\pi }
\int _{0} ^{2\pi} \frac{d \varphi }{2\pi }
\int d \psi ^{+ }d \psi    $.
${\scriptstyle \chi } $ indicates  that
Grassmann numbers will  acquire a minus sign.
In this solution we have used a so-called
reduced measure (see \cite {Weller,et}).
There is another  resolution and a trace with  the
invariant measure
$\int= \frac{1}{2\pi ^{2}}
\int _{0} ^{\pi /2}d \theta \sin (2\theta )
\int _{0} ^{2\pi} d \phi
\int _{0} ^{\pi /2} d \vartheta \sin (2\vartheta )
\int _{0} ^{2\pi} d \varphi
\int d \psi^{\dagger}  d \psi  $;
however the Lagrangian will be independent
of this choice.  \\
The expectation values of the Hubbard operators are listed in the following
table.\\

\begin {tabular}{|l|l|} \hline \hline
\rule[-1mm]{0cm}{6mm} bosonic &  fermionic\\ \hline \hline
\rule[0mm]{0cm}{4mm}$  \langle G |  X^{00}| G \rangle  = \psi ^{\dagger} \psi
\cos ^{2} \left( \theta \right)  $&
$
\langle G |  X^{\+ 0}| G \rangle  =\psi \sin \left(\vartheta \right)e
^{-i \varphi }\cos \left( \theta\right) $\\
$\langle G |  X^{\+ \+ } \  | G \rangle  =\left( 1+ \psi \psi ^{\dagger}
\right) \sin ^{2} \left( \vartheta \right)$ &
$
\langle G |  X^{0\+ }| G \rangle  =\psi ^{\dagger}\sin \left(\vartheta \right)e
^{i \varphi }\cos \left( \theta\right)  $\\
$\langle G |  X^{\- \- } \  | G \rangle  =\left( 1+ \psi \psi ^{\dagger}
\right) \cos ^{2} \left( \vartheta \right)$&
$
\langle G |  X^{\- 0}| G \rangle  =\psi \cos \left( \vartheta \right)
\cos \left( \theta\right) $\\
$\langle G |  X^{22}| G \rangle  = \psi ^{\dagger} \psi \sin ^{2} \left( \theta
\right)  $ &
$
\langle G |  X^{0\- }| G \rangle  =\psi ^{\dagger} \cos \left( \vartheta
\right)\cos \left( \theta \right)
 $ \\
$\langle G |  X^{02}| G \rangle  =\psi ^{\dagger} \psi
\sin \left( \theta \right)\cos \left(
\theta \right)e ^{i \phi }  $&
$
\langle G |  X^{\+ 2}| G \rangle  =\psi \sin \left(\vartheta \right)e
^{-i \varphi }
\sin \left( \theta\right)e ^{i \phi } $\\
$
\langle G |  X^{20}| G \rangle  =\psi ^{\dagger} \psi
\sin \left( \theta \right)\cos \left(
\theta \right)e ^{-i \phi }  $&
$
\langle G |  X^{2\+ } | G \rangle  =\psi ^{\dagger} \sin \left(\vartheta
\right)e ^{i \varphi }
\sin \left( \theta \right)e ^{-i \phi } $ \\
$\langle G |  X^{\+ \- } \  | G \rangle  =\left( 1+ \psi \psi ^{\dagger}
\right)
\sin \left( \vartheta \right)\cos \left(
\vartheta \right)e ^{-i \varphi } $ &
$
\langle G |  X^{\- 2}| G \rangle  =\psi \cos \left( \vartheta \right)\sin
\left( \theta\right)e ^{i \phi } $\\
$\langle G |  X^{\- \+ } \  | G \rangle  =\left( 1+ \psi \psi ^{\dagger}
\right)
\sin \left( \vartheta \right)\cos \left(
\vartheta \right)e ^{i \varphi }   $&
$
\langle G |  X^{2\- }| G \rangle  =\psi ^{\dagger} \cos \left( \vartheta
\right)\sin \left( \theta \right)e ^{-i
\phi }  $
\\ \hline \hline
\end {tabular}
\\

This table allows  $(\vartheta, \varphi )$
to be interpreted as   angular spin variables, and
$(\theta, \phi )$ as angular charge variables.
The charge variables are pseudospin variables in the space of
empty and doubly occupied states.
They parameterize rotations in this space, as $(\vartheta, \varphi )$ do in the
spin space. The Grassmann field induces transitions between spin and pseudospin
space,
e.g.\ $\psi ^{\dagger }$ creates a hole in the constrained space and a doubly
occupied
state in  the complementary space.

\section{Path Integral Representation}
\label{app4}

In the previous section  we introduced coherent states for the Hubbard
operators
and fixed the measure which guarantees a resolution of the unity.
Now, using the Trotter formula and
this   resolution of the  unity operator with  the  coherent state
(\ref{kohZust} ),
we are in the position  to  set up  a path integral for the partition function.
The derivation is   analogous to the spin $1/2$ case in appendix \ref{Anh1}.
\begin{eqnarray}
Z =  \int {\cal D } \left[ \theta ( \tau ),
\phi (\tau ), \vartheta ( \tau ),
\varphi (\tau ), \psi ^{\dagger}(\tau),\psi (\tau) \right]
\exp{\left[ -\int _{0}^{\beta }\left(
L_{0} +L_{1}+L_{t} \right) d \tau \right] } \label{pathint}
\end{eqnarray}
where the  Berry phase $L_{0}$ is:
\begin{eqnarray}
L_{0}=\langle G(\tau )|\partial _{\tau }|G(\tau )\rangle =
\sum _{i} \left[ i\dot{\varphi _{i}}\sin^{2}{\vartheta _{i}} +
\psi^{\dagger} _{i}\left( \partial _{\tau }
+ i \dot{\phi _{i}}\sin^{2} {\theta _{i}}
-i \dot{\varphi _{i}}\sin^{2} {\vartheta _{i}}
\right)
\psi _{i} \right]
\end{eqnarray}
The Hamiltonian part may be directly
taken from the table. It contains the  Coulomb potential  $UX^{22}$ and
the chemical potential $\mu \left(\sum _{\sigma } X^{\sigma \sigma
}+2X^{22}\right) $ which corresponds to
\begin{eqnarray}
L_{1}= \sum _{i} \left[
-\mu + \psi^{\dagger}_{i}\left(  \frac{U}{2} + \left( \mu -\frac{U}{2} \right)
\cos (2\theta _{i}) \right) \psi _{i}\right]
\end{eqnarray}
The form of $L_{1}$ allows to interpret $\mu -U/2$ as a magnetic field
coupling to the pseudospin (c.f.\ appendix \ref{Anh1}).
The discrete  version of the local Lagrangian
$L_{0} + L_{1}$, will be presented in the next section.

Further it contains  the kinetic  term which
we write in the discrete version in order to
exhibit the correct time ordering of the field variables:
\begin{eqnarray}
L_{t}= L_{t;01}+L_{t;20}+L_{t;11}+L_{t;21}
\end{eqnarray}
where
\begin{eqnarray}
L_{t;01}&=&
\sum_{<i,j>} t_{ij}   \psi_{i,n-1} \psi_{j,n}^{\dagger}
\cos{\theta_{i,n-1}}
\cos{\theta_{j,n}}
\nonumber \\
&& \times \left(\sin{\vartheta_{i,n}}\sin{\vartheta_{j,n-1}}
e^{-i\left( \varphi _{i,n}-\varphi _{j,n-1}\right)}
+ \cos{\vartheta_{i,n}} \cos{\vartheta_{j,n-1}}\right)\label{Lt}
\end{eqnarray}
represents the exchange of a singly occupied site with its empty
nearest neighbor,
\begin{eqnarray}
L_{t;20}&=&\sum_{<i,j>} t_{ij}
\psi _{i,n-1} \psi _{j,n-1}
\cos{\theta _{i,n-1}}
\sin{\theta _{j,n-1}} e^{i\phi _{j,n-1}}
\nonumber \\
&& \times
\left( \sin{\vartheta _{i,n}} \cos{\vartheta _{j,n}}
e^{-i \varphi _{i,n}}
- \cos{\vartheta _{i,n}} \sin{\vartheta _{j,n}}e^{i \varphi _{j,n}} \right)
\end{eqnarray}
signifies  a   transition from a state with  two singly
occupied nearest neighbor  sites to a state with neighboring
doubly occupied and   empty sites,
\begin{eqnarray}
L_{t;11}&=&\sum_{<i,j>} t_{ij} \psi_{i,n}^{\dagger} \psi_{j,n} ^{\dagger}
\sin{\theta_{i,n}}
\cos{\theta_{j,n}}e^{-i\phi _{i,n}}
\nonumber \\
&& \times
\left(\cos{\vartheta_{i,n-1}}\sin{\vartheta_{j,n-1}}
e^{i \varphi _{j,n-1}}
- \sin{\vartheta_{i,n-1}} \cos{\vartheta_{j,n-1}}e^{-i \varphi _{i,n-1}}
\right)
\end{eqnarray}
is the inverse process, and
\begin{eqnarray}
L_{t;21}&=&\sum_{<i,j>} t_{ij}
\psi_{i,n}^{\dagger} \psi_{j,n-1}
\sin{\theta_{i,n}}
\sin{\theta_{j,n-1}}e^{-i(\phi _{i,n}-\phi _{j,n-1 })}
\nonumber \\
&& \times
\left( \cos{\vartheta_{i,n-1}} \cos{\vartheta_{j,n}}
+ \sin{\vartheta_{i,n-1}} \sin{\vartheta_{j,n}} e^{i\left(
       \varphi _{i,n-1} -  \varphi _{j,n}            \right) }
       \right)
\end{eqnarray}
denotes
the exchange of a singly occupied site with its doubly
occupied nearest neighbor.
We check that, if   we set   all  angle-variable fields to zero, we  find
a Lagrangian  which describes spinless fermions  up to a
particle hole transformation (cf.\  appendix \ref{Anhang5}).

\section{Atomic Limit }
\label{att}

\subsection{Continuum Version}
\label{Cont}

The
partition function may be calculated easily in the case of no band motion:
$t_{ij}=0$.  Although this atomic limit is trivial physicswise, we consider it
carefully in order to test the integration techniques and further to develop
projection
methods in the next two sections.
In  this atomic limit we integrate out the Grassmann field using   appendix
\ref{Anhang2}.
\begin{eqnarray}
Z_{at}&=& \int {\cal D } \left[ \theta ( \tau ),
\phi (\tau ), \vartheta ( \tau ),
\varphi (\tau ), \psi ^{\dagger}(\tau), \psi (\tau) \right] \ \
\exp \left[
-\int _{0} ^{\beta } d \tau \ \left(
L_{0} + L_{1}
\right) \right]       \nonumber  \\
&=& \int {\cal D }\left[ \vartheta ( \tau ),
\varphi (\tau )  \right]
\exp \left[
\mu \beta -\int  _{0} ^{\beta }d \tau \ i\dot{\varphi }\sin^{2}{\vartheta
}\right]
\int {\cal D }\left[ \theta ( \tau ),
\phi (\tau ) \right]   \\
 & &  +
\int {\cal D }\left[ \theta ( \tau ),
\phi (\tau ) \right]
\exp \left[  -\int  _{0} ^{\beta } d \tau \
\left( i\dot{\phi }\sin^{2}{\theta }
+ \left( \mu -\frac{U}{2} \right)
\left( \cos(2\theta ) -1\right) \right)  \right]  \int {\cal D }\left[
\vartheta ( \tau ),
\varphi (\tau )  \right]    \nonumber
\end{eqnarray}
The first term is a path integral for the free spin $1/2$ and, therefore,
yields
$2e^{\mu \beta }$. The second term is also a spin path
integral, but
with the pseudo  angle variables  and
subject to
a `magnetic field'
$\mu - U/2$. Therefore we  obtain
\begin{eqnarray}
Z=2e^{\mu \beta } + \left( 1 + e^{(2\mu -U )\beta } \right)
\end{eqnarray}

\subsection{Discrete Version}
\label{AusintatLim}

As is well known, the continuum version of the action is only a short form,
correct
to first order in $\beta /N$, but
several situations necessitate to consider the second order, e.g.\
magnetic field problems \cite{Schulmann} or  particle
hole transformations (appendix \ref{Anhang5}).
Therefore we reformulate the previous subsection in the discrete language
and introduce some notation which will be used in the following.

The vector notation of spin coherent states   at   time step $l$ is:
\begin{eqnarray}
|{\bf n }_{l} \rangle = \left(
\begin{array}{l}
\sin (\vartheta _{l} )e^{i\varphi _{l}}\\
\cos (\vartheta _{l} )
\end{array}\right)
\end{eqnarray}
and for the pseudo spin coherent  state:
\begin{eqnarray}
|{\bf N }_{l}  \rangle = \left(
\begin{array}{l}
\sin (\theta _{l} )e^{i\phi _{l}}\\
\cos (\theta _{l} )
\end{array}\right)
\end{eqnarray}
To   first order in  $\epsilon =\beta /N$ we  are allowed to use:
\begin{eqnarray}
1-\langle {\bf n }_{l}  |  {\bf n }_{l-1 }\rangle  =
i\dot{\varphi }\sin^{2}{\vartheta } \epsilon + {\cal O (\epsilon )}
&\qquad\qquad & \langle {\bf n }_{l}  |\sigma _{z} |{\bf n }_{l-1 }  \rangle =
-\cos (2 \vartheta _{l}) + {\cal O (\epsilon )}\label{formel1} \\
1-\langle {\bf N }_{l}  |  {\bf N }_{l-1 }\rangle  =
i\dot{\phi }\sin^{2}{\theta } \epsilon + {\cal O (\epsilon )}
&\qquad\qquad & \langle {\bf N }_{l}  |\sigma _{z} |{\bf N }_{l-1 }  \rangle =
-\cos (2 \theta _{l}) + {\cal O (\epsilon )}\label{formel2}
\end{eqnarray}
With these relations we rewrite the atomic limit in the following form:
\begin{eqnarray}
Z_{at}&=& \int \left( \prod _{l=1}^{N}\frac{1}{2 \pi ^{4} }
d \vartheta _{l} d \varphi _{l} d \theta
_{l}
d \phi_{l}  d
\psi _{l} d \psi ^{\dagger}_{l}  \right)\, e^{\mu \beta }
\prod _{l=1}^{N}\langle {\bf n }_{l}  |  {\bf n }_{l-1} \rangle  \\
&&\times \exp \left[ - \sum _{l=1}^{N} \left(
\psi ^{\dagger}_{l} \psi_{l}
+ \psi ^{\dagger}_{l} \psi _{l-1}
\left( -1+
\langle {\bf n }_{l}  | {\bf n }_{l-1}  \rangle  -
\langle {\bf N }_{l}   |
\1 - \frac{U}{2} \1 \epsilon
+(\mu - \frac{U}{2})\sigma  _{z} \epsilon
|{\bf N }_{l-1} \rangle
\right) \right)
\right] \nonumber
\end{eqnarray}
Expanding the exponential function and exploiting the fact that
all higher orders
vanish because of the anticommuting properties of the Grassmann
fields $\psi , \psi ^{\dagger}$ leads to\footnote{In functional integrals over
Grassmann fields it is necessary to include the zeroth
time slice explicitly due to the antiperiodic boundary conditions. However, in
order to present
the following calculations in a more readable form we will always skip
this term.}:
\begin{eqnarray}
Z_{at}&=& \int \left( \prod _{l=1}^{N}\frac{1}{2\pi ^{4} }d \vartheta _{l} d
\varphi _{l} d \theta
_{l}
d \phi_{l}  d
\psi _{l} d \psi ^{\dagger}_{l}  \right)\,
e^{\mu \beta }
\prod _{l=1}^{N}\langle {\bf n }_{l}  |  {\bf n }_{l-1} \rangle
\nonumber    \\
& &
\times \prod _{l=1}^{N}\left(
1+\psi _{l} \psi ^{\dagger}_{l} +
\psi ^{\dagger}_{l} \psi _{l-1}
\left( 1-
\langle {\bf n }_{l}  | {\bf n }_{l-1}  \rangle  +
\langle {\bf N }_{l}   |
\1 - \frac{U}{2} \1 \epsilon
+(\mu - \frac{U}{2})\sigma  _{z} \epsilon
|{\bf N }_{l-1} \rangle
\right)
\right)\nonumber
\end{eqnarray}
In order to carry out the integrations   we  first   collect
the complete Grassmann chains as explained
in   appendix \ref{Anhang2}.
\begin{eqnarray}
Z_{at}
&=& \int \left( \prod _{l=1}^{N} \frac{1}{2\pi ^{4} }d \vartheta _{l}
d \varphi _{l} d \theta
_{l}
d \phi_{l}  d\psi _{l} d \psi ^{\dagger}_{l}  \right) \ e^{\mu \beta }
\left[ \prod _{l=1}^{N} \left( \psi _{l} \psi ^{\dagger}_{l}\langle {\bf n
}_{l}
 | {\bf n }_{l-1}  \rangle
\right) \right. \nonumber \\ & & \ \ \left. \ \ \ \ \ \ +
\prod _{l=1}^{N}\left( \psi ^{\dagger}_{l} \psi _{l-1}
\langle {\bf N }_{l}  |
\1 - \frac{U}{2} \1 \epsilon
+(\mu - \frac{U}{2})\sigma  _{z} \epsilon
|{\bf N }_{l-1}  \rangle \right)
\right]
\end{eqnarray}
Integration over the Grassmann and spin (first expression) or
pseudo spin (second expression) variables yields:
\begin{eqnarray}
Z_{at} &=&
\int \left( \prod _{l=1}^{N}\frac{1}{ \pi ^{2} }d \vartheta _{l} d
\varphi _{l}
  \right) \
\prod _{l=1}^{N} \langle {\bf n }_{l}  | {\bf n }_{l-1}  \rangle e^{\mu \beta }
\nonumber \\
& & \ \ \ +
\int \left( \prod _{l=1}^{N}\frac{1}{ \pi ^{2} } d \theta  _{l}
d \phi _{l}   \right) \,
e^{\mu \beta }
\prod _{l=1}^{N}
\langle {\bf N }_{l}  |
\1 - \frac{U}{2} \1 \epsilon
+(\mu - \frac{U}{2})\sigma  _{z} \epsilon
|{\bf N }_{l-1}  \rangle
\end{eqnarray}
Using the resolution of the unity operator we immediately obtain:
\begin{eqnarray}
Z_{at}    &=&
\Sp (\1 )e^{\mu \beta }+e^{\mu \beta }
\Sp \left[
 \exp \left[ -\frac{U}{2}\1 \beta
+(\mu -\frac{U}{2})\sigma _{z} \beta  \right]
\right] \nonumber \\
&=&
2e^{\mu \beta } + 1 + e^{(2\mu -U )\beta }\label{atlimausgr}
\end{eqnarray}

\section{Methods of Projection Using  Spin Coherent States }
\label{pro}

In this section we introduce several methods of projection with
the   intention    of  applying  one of these   to  the  path integral
(\ref{pathint}).
It was shown in  section \ref{app4} that the   local Coulomb interaction $U$
is similar to a magnetic field in the pseudo spin space.
Therefore,  we first  deal with the
system of a `spin in a magnetic field' and investigate     the limit of an
infinitely  strong magnetic  field. This  will result in a procedure
with a well-defined measure which
allows to take the limit
of strong interaction $U\rightarrow \infty$.

\subsection{Method 1: Direct Integration }

We write the spin path integral in  its discrete version
\begin{eqnarray}
Z=
\lim _{N\rightarrow \infty } \left( \frac{1}{\pi ^{2} }\right) ^{N}
\int  \left( \prod _{n=1 }^{N}
d \varphi _{n} d \vartheta  _{n} \right)
\exp \left[ - i \sum _{n=1} ^{N}
\left(
(\varphi _{n} - \varphi _{n-1} )
\sin ^{2}(\vartheta _{n})
\right)
\right] \label{gl11}
\end{eqnarray}
and carry out the    angular integrations. To this end, the $\varphi $
integration is extended  to
$L$ intervals and
$\sin ^{2}(\vartheta _{n})=\frac{1}{2}\left( 1-\cos(2 \vartheta _{n})\right)$
is replaced.
We find:
\begin{eqnarray}
Z\! & =& \!
\lim _{
N\rightarrow \infty} \lim _{ L\rightarrow \infty }
\left( \frac{1}{\pi ^{2} } \right) ^{N}
\left(  \prod _{n=1 }^{N}  \int _{-2\pi L } ^{2\pi L}
\frac{d \varphi _{n}}{2L} \int _{0} ^{\pi /2 } d \vartheta  _{n}
\right)
\nonumber \\ & & \times
\exp \left[  i \sum _{n=1} ^{N}
(\varphi _{n} - \varphi _{n-1} ) \frac{1}{2}
\cos (2 \vartheta _{n})
\right] \label{ddddd}
\end{eqnarray}
where   periodic boundary conditions were used. Integrating out the
$N$ $\vartheta $ variables,   a $N$-fold
convolution product of Bessel functions $J_{0}$ emerges which
may be diagonalized by Fourier transformation.
Furthermore  we   integrate out the
$\varphi _{n} $ and use the resulting $\delta $ functions.
It follows:
\begin{eqnarray}
Z=
\lim _{N\rightarrow \infty } c \cdot
\int _{-1/2}^{1/2} \frac{d k}{2 \pi } \
\left(
\frac{1}{\sqrt{\frac{1}{4}- k ^{2}}}
\right) ^{N}
\end{eqnarray}
where $c=(1/\pi)^{N}$.
Taking the limit $N\rightarrow \infty$, the only surviving contributions
 are  due to  both poles at $k=\pm 1/2 $.
In order to determine the contributions, we introduce a cutoff  $\epsilon _{c}
$.
Integration over  $k$ yields:
\begin{eqnarray}
Z=2\lim _{N\rightarrow \infty }
\lim _{\epsilon _{c}\rightarrow 0  }\frac{1}{\epsilon _{c}\pi (N+2)}
\left( \frac{2}{\sqrt{\epsilon _{c}}}\right) ^{N} \label{cons}
\end{eqnarray}
The prefactor 2 arises from  the two poles. To solve the problem of the
continuum limit
in the spirit of Feynman requires only the quotient of  path integrals
with different couplings to be well-defined.  The
existence of this quotient is guaranteed by the precise knowledge
of the normalization factor.
In our case the limiting procedure is explicitly known (see (\ref{cons})).
Specifically, the problem of a spin in a `magnetic field'
is solved analogously to (\ref{gl11})--(\ref{cons}) and yields the same
normalization factor as in
(\ref{cons})---independent of the `magnetic field'.
Therefore the existence of   quotients such as  $Z(B)/Z(B=0)$ is ensured.

\subsection{Method 2: Time Dependent Saddle Point Approximation  }

The spin path integral is the sum over all paths   on a sphere.
A path is a set containing   points
for each time and space step.
In order to map the surface of the sphere onto planar coordinates
uniquely the sphere is cut along the equator and
each of  the hemispheres is represented  by a separate plane.
With this stereographic projection each point will be placed on either
of the two planes.

There is a natural way to extend the standard saddle point approximation
to this spin path integral:
at  every
space and time step the functional  will be expanded around
a point in each of the two planes.
In this section we will clarify this procedure and show that it results
in the exact expression  for the partition function. \\

For the explicit evaluation of the partition function of a single spin, first,
we
chose north and south pole as reference points  for the expansion:
\begin{eqnarray}
Z&=&
\lim _{N\rightarrow \infty } \left( \frac{1}{ \pi ^{2} }\right) ^{N}
\int  \left( \prod _{n=1 }^{N}
d \varphi _{n} d \vartheta  _{n} \right)
\exp \left[  i \sum _{n=1} ^{N}
\left(
(\varphi _{n} - \varphi _{n-1} ) \frac{1}{2}
\cos (2\vartheta _{n})
\right)
\right]
\label{zitierespin} \\
&=& \lim _{N \rightarrow \infty }
\left( \frac{1}{ \pi ^{2} }\right) ^{N} \int \prod _{n=1 }^{N} \left[
d{ \varphi } _{n} d \tilde{ \vartheta } _{n} \right]
\prod _{n=1 }^{N} \left[
e^{-i( { \varphi } _{n}- { \varphi }
_{n-1})\left( \frac{1}{2} -\tilde{ \vartheta   }
_{n} ^{2} \right) }  +
e^{+i( { \varphi }_{n}- { \varphi }
_{n-1})\left( \frac{1}{2} -\tilde{ \vartheta  }
_{n} ^{2} \right) }
\right] \nonumber
\end{eqnarray}
where we have expanded the exponentials  up to   second order in $
 \tilde{ \vartheta } _{n} $, where
the field $
 \tilde{ \vartheta } _{n} $ is  the fluctuation
around   north and south pole.
Next we integrate out the
$ \tilde{ \vartheta }_{n} $. The integration runs from 0
to  arbitrary $\rho $.
The result is an infinite convolution of Fresnel functions  which can
be diagonalized by a Fourier transformation.
Making use of the periodic boundary conditions for $\varphi _{n}$
the partition function reads:
\begin{eqnarray}
Z&=&   \lim _{N \rightarrow \infty }c \cdot \int d k \
\left[
\frac{1}{\sqrt{k+\frac{1}{2} } } +\frac{1}{\sqrt{k-\frac{1}{2} } }
\right] ^{N}
\end{eqnarray}
The only contributions  to  the integral  are due to
the poles at $\pm \frac{1}{2}$. \\

One might worry that $c$ has to be tuned separately
for each kind of spin system.
Hence one might infer that the quotient of two
path integrals is not well-defined.
Therefore we   calculate  two spins in a magnetic field
interacting via  a Heisenberg coupling.
Using the above saddle point approximation and introducing auxiliary fields,
to write $\cos (\vartheta _{n})$ as a derivative to
$ \varphi _{n}-\varphi  _{n-1}$, we  show that the
same  limiting procedure, using  $c$ and cutoff $\epsilon_{c}$,
normalizes this more complicated problem.\\

First, we present the calculation for  an Ising coupling  of two
spins in the presence of  a
magnetic field.  The Lagrangian is:
\begin{eqnarray}
L=-\sum _{i=1,2} \frac{ix_{n,i}}{2\epsilon }\cos (2 \vartheta _{n,i})
+\frac{J}{4} \cos (2\vartheta _{n,1} )
\cos (2\vartheta _{n,2} ) +B \sum _{i=1,2} \cos (2 \vartheta _{n,i})
\label{sdew}
\end{eqnarray}
where $x_{n,i}=\varphi _{n} - \varphi _{n-1}$.
Now we work with the Gaussian method discussed  above.
Therefore, we have to add up the four combinations of north and south pole
at each site.
Rewriting the  Ising coupling and
magnetic field    with auxiliary fields, the partition function reads:
\begin{eqnarray}
&&Z=\left( \frac{1}{\pi ^{2}L}\right) ^{2N}\int \prod _{i=1,2}
 \prod _{n=1} ^{N} d\varphi _{n,i}d \vartheta _{n,i} \
 \Bigg\{ \sum _{  \sigma _{1}    =\pm 1}\sum _{  \sigma _{2}    =\pm 1}
\nonumber \\ & & \! \! \! \! \! \! \! \!
\times \exp \left[  iJ/2 \epsilon (\sigma _{1}\partial _{x_{n,1}}+
\sigma _{2}\partial _{x_{n,2}}+
\sigma _{1}\sigma _{2}\frac{i}{2})
+2B\epsilon i(\partial _{x_{n,1}}+\partial _{x_{n,2}}) \right] \nonumber \\ & &
\times \exp \left[ \sigma _{2}ix_{n,1} \left( \frac{1}{2} -\vartheta ^{2}
_{n,1}
\right)   +\sigma _{1}ix_{n,2} \left( \frac{1}{2} -\vartheta ^{2} _{n,2}
\right)  \right]\Bigg\}
\end{eqnarray}
Next we have to do the $\vartheta $ integrations  which cover the
interval from 0  to $\rho$.  Fourier transformation and straightforward
calculation yields:
\begin{eqnarray}
Z =  \lim _{N\rightarrow \infty} c^{2} \cdot \int _{-\infty}^{\infty } dk_{1}
\int _{-\infty}^{\infty } dk_{2} \left( f(k_{1},k_{2}) \right) ^{N}
\end{eqnarray}
where
\begin{eqnarray}
f(k_{1},k_{2})&=&\frac{1}{2}\sum _{\sigma_{1},\sigma_{2}}\left[
e^{\sigma _{1}J\epsilon /4} + e^{-\sigma _{1}J\epsilon /4}\left(
\cos (iJ\epsilon /2) -2i\sigma _{1}\sin(iJ\epsilon /2)\right) \right] \times
\nonumber \\ && \times
\frac{e^{-2B\epsilon }}{ \sqrt{ 1/2 +\sigma _{2}k_{1}  }\sqrt{1/2 +
\sigma _{1}\sigma _{2}k_{2} }   }
\end{eqnarray}
The dependence   on $\rho $ has vanished  because we have applied
formulas of the following kind:
\begin{eqnarray}
\int _{0} ^{\infty } \frac{dx}{2\rho }
\frac{S(\sqrt{x})}{\sqrt{x}}
\sin (\frac{b^{2}x}{\rho ^{2}})
=\frac{\sqrt{\pi}2^{-5/2}}{b}
\end{eqnarray}
Using the same cutoff and normalization constant as in method 1, we obtain:
\begin{eqnarray}
Z=2e^{J\beta /2} \cosh(B\beta ) +2e^{J\beta /2}
\end{eqnarray}
which is the correct result for the model (\ref{sdew}).
Extending these considerations to a Heisenberg coupling,   we  switch on the
$x-y$ interaction, expand in the $x-y$ coupling   to   all orders,
calculate  the value of an arbitrary term,  and sum  up all  terms
to  leading order  in  $\epsilon$.
Using the same limiting procedure and  normalization,
we  find  the correct result for the partition function \cite{et}. \\

Later on, we will need a well-defined projection method to handle the
$U\rightarrow \infty$ limit.
Thereby we note that the north pole collects a weight $e^{\beta B}$
and the south pole  $e^{-\beta B}$.
The only surviving path in the limit of an infinitely strong magnetic field
is the path  that  stays at the north pole over the full  time range.
However,  if we set $\vartheta =0$,    we would not  find the same
normalization constant  as before.
The correct  way of doing the projection  is the following:
\begin{eqnarray}
L=i\dot{ \varphi } \sin ^{2}(\vartheta ) + B \cos (2 \vartheta  )
\rightarrow L=i\dot{ \varphi } \sin ^{2}(\vartheta )\Big|
_{\mbox{\scriptsize north pole}} + B
\label{prescription}\end{eqnarray}
In this prescription the Berry phase is not eliminated,  but has some
small fluctuations around the pole.
In the Hamiltonian part,  the prescription yields  the same result, as
if we  had  set $\vartheta =0$ without further considerations.
We note that this prescription does not change the measure.

In the discrete language the corresponding prescription is:
\begin{eqnarray}
&&\int \left( \prod _{l=1}^{N} \frac{1}{\pi ^{2} } d \vartheta  _{l}
d \varphi _{l}   \right)
\prod _{l=1}^{N}
\langle {\bf n}_{l}  | \left(
\1
+B\sigma _{z} \epsilon \right)
|{\bf n}_{l-1}  \rangle \nonumber \\
&&\rightarrow
\int \left( \prod _{l=1}^{N} \frac{1}{\pi ^{2} } d \vartheta  _{l}
d \varphi _{l}   \right)
\prod _{l=1}^{N}
\langle {\bf n}_{l}  | \left(
\1
+B\sigma _{z} \epsilon \right) P_{\mbox{\scriptsize north pole}}
|{\bf n}_{l-1}  \rangle \label{pre}
\end{eqnarray}
where $P_{\mbox{\scriptsize north pole}} =\left(
\begin{array}{cc} 0 &0  \\  0 & 1
\end{array} \right) $.
The path integral for infinite `magnetic field'
may be either gained by using (\ref{prescription}) in the continuum version
(analogously to \ref{Cont}) or by using
(\ref{pre}) in the discrete version (analogously to \ref{AusintatLim}).
As compared to section \ref{att}, the projection onto the constrained space is
now
implemented into the path integral formalism ab initio.

\subsection{Method 3: Action Angular Variables }

In this section we want to present a method, interesting in a didactical sense.
First, we consider the Berry phase of the spin $1/2 $ path integral.
\begin{eqnarray}
Z=\lim _{N \rightarrow \infty }\left( \frac{1}{ \pi ^{2} }\right)
^{N} \int \left(
\prod _{n=1}^{N}
d \vartheta _{n} d \varphi _{n }\right) \exp \left[
\frac{i}{2} \sum_{n=1 }^{N}
\left( \varphi _{n} -\varphi _{n-1}\right) \cos  (2\vartheta _{n})
\right]
\end{eqnarray}
There is a geometrical interpretation  to  this expression:
The Berry phase is the area enclosed by a single  path.
This reminds  us of classical action angle variables in
classical mechanics.
Therefore, we want to  implement similar variables.\\

To do so, we start by separating   the partition function into a  sum
over  all  right
and left rotating paths:
 \begin{eqnarray}
Z&=&
\frac{1}{2}\lim _{N \rightarrow \infty }\left( \frac{1}{ \pi ^{2} }\right)
^{N} \int
\left( \prod _{n=1}^{N}
 d \vartheta _{n} d \varphi _{n }\right)
\exp \left[ -\frac{i}{2} \sum_{n=1 }^{N}
\left( \varphi _{n} -\varphi _{n-1}\right) \cos  (2\vartheta _{n})
 \right]\nonumber \\
& & +
\frac{1}{2}\lim _{N \rightarrow \infty }\left( \frac{1}{ \pi ^{2} }\right)
^{N} \int  \left(
\prod _{n=1}^{N}
 d \vartheta _{n} d \varphi _{n }\right)
\exp \left[ \frac{i}{2} \sum_{n=1 }^{N}
\left( \varphi _{n} -\varphi _{n-1}\right) \cos  (2\vartheta _{n})
 \right]\nonumber \\
&=&
\frac{1}{2} \left( Z_{1}+Z_{1}^{\star } \right)
\end{eqnarray}
The classical action variable is the area enclosed by a single path.
Therefore the action  of the single spin is just this classical action
variable,
and it seems to be advantageous to transform  the $\vartheta $ field variables
into   action variables.
Since the field $\vartheta $ consists of $N$ $\vartheta _{n}$-variables
(and there
exists only one classical action variable), we introduce new variables
being defined as   the areas which are swept out by the geodesic line from
the north pole to the actual position on the considered path (see figure).
In this way  we obtain $N$ variables:
\begin{eqnarray}
B_{n}= \frac{1}{2} \sum_{l=1 }^{n}
\left( \varphi _{l} -\varphi _{l-1}\right) \cos  (2\vartheta _{l})
\end{eqnarray}
Then $B_{N}$ is  the  Berry phase.
The Jacobian is:
\begin{eqnarray}
|J|^{-1}=\left|
\begin{array}{cc}
\frac{\partial B }{\partial \vartheta } &
\frac{\partial B }{\partial \varphi }  \\
\frac{\partial \varphi  }{\partial \vartheta  }  &
\frac{\partial \varphi  }{\partial \varphi  }
\end{array}
\right|
=
\left|
\begin{array}{cc}
\frac{\partial B }{\partial \vartheta } &
\frac{\partial B }{\partial \varphi }  \\
0 & \1
\end{array}
\right|
\end{eqnarray}
where
\begin{eqnarray}
\frac{\partial B _{n} }{\partial \vartheta _{m} }
=
\left\{
\begin{array}{ll}
(\varphi _{m-1}-\varphi _{m}) \sin (2 \vartheta _{m}) & m \leq n \\
0 & m \ge n
\end{array}
\right.
\end{eqnarray}
is a triangular matrix and we  find explicitly:
\begin{eqnarray}
|J|^{-1}= \prod _{l=1}^{N}
\left|
(\varphi _{l}-\varphi _{l-1}) \sin (2 \vartheta _{l})
\right| =  \sqrt{(\varphi _{1}-\varphi _{0})^{2}  + 4 B_{1} ^{2} }
\ \prod _{l=2} ^{N}\sqrt{(\varphi _{l}-\varphi _{l-1})^{2} + 4
(B_{l} - B _{l-1})^{2} }\nonumber
\end{eqnarray}
The partition function  reads:
\begin{eqnarray}
Z = \lim _{N \rightarrow \infty }\left( \frac{1}{ \pi ^{2} }\right)
^{N} \int \left(
\prod _{n=1}^{N}
 d B _{n} d \varphi _{n } \right)
|J|^{-1}
e^{iB_{N} },
\end{eqnarray}
The action  has  become trivial  but the integration over  the Jacobian
is   a $N$-fold convolution.
Fourier transformation decomposes the convolution into a product of independent
integrals.
\begin{eqnarray}
\frac{1}{2}
Z_{1}&=& \int d B_{N} d \varphi _{N} d k _{1,N}
d k_{2,N}\  e^{i B_{N}+i B_{N}k _{2,N} } \
f^{N} (k _{1,N} , k _{2,N}  )\nonumber \\
&=&\int d k_{1,N}\ f^{N}(k_{1,N}, -1)
\end{eqnarray}
and
\begin{eqnarray}
\frac{1}{2}
Z_{1}^{*}=\int d k_{1,N}\ f^{N}(k_{1,N}, +1)
\end{eqnarray}
where
\begin{eqnarray}
f( k _{1}, k _{2}) &=& \int _{xy} \left[ x^{2} + y^{2} \right] ^{-1/2}
\exp \left[ ik _{1}x+ k _{2}y  \right] \nonumber \\
&=&
\int r d r d \varphi \
\frac{e^{ikr\cos \varphi }}{r}\nonumber \\
&\sim & \frac{1}{k} = \frac{1}{\sqrt{ k _{1}^{2}+  k _{2}^{2} }}
\end{eqnarray}
The normalization of the resulting singularities is similar to the procedure
of the section above (see \ref{ddddd}-\ref{cons}).
Furthermore, the analogous scheme applied to a spin in a magnetic field
yields the correct result.

\section{Alternative Projection Method Using  Fermionic Coherent States}
\label{erer}

In  the first part of this section  we  present an  alternative projection
method
using the standard path integral formalism with
two  Grassmann fields.
The basic idea is to perform projections at each time step explicitly through
a suitable choice of the measure.
Then, we introduce coherent states in the constrained space and show that their
measure is identical to the `projecting' measure of the first part.  \\
First, we repeat the calculation for a local action, e.g.\
a chemical potential term:
\begin{eqnarray}
Z_{at} &=& \int \left( \prod _{n=1} ^{N}
d\psi _{\+ n}^{\dagger}d\psi _{\+ n}
d\psi _{\-n}^{\dagger}d\psi _{\-n}
\right) \
\exp \left[ -\sum _{n=1 }^{N} \sum _{\sigma =\left\{ \+ , \-
\right\} }
\left( \psi ^{\dagger} _{\sigma , n} \psi _{\sigma ,n}-
\psi ^{\dagger} _{\sigma ,n} \psi _{\sigma ,n-1} (1+\mu \epsilon )
\right) \right] \nonumber \\
&=&   \int \left( \prod _{n=1} ^{N}
d\psi _{\+ n}^{\dagger}d\psi _{\+ n}
d\psi _{\-n}^{\dagger}d\psi _{\-n}   \right) \ \nonumber
\\ & &
\times \Bigg\{
\left( \prod _{n=1} ^{N} \psi _{\+ ,n} \psi ^{\dagger} _{\+ ,n} \right)
\left[
\left( \prod _{n=1} ^{N} \psi _{\-,n} \psi ^{\dagger} _{\-,n} \right)
+\left( \prod _{n=1} ^{N} \psi ^{\dagger} _{\- ,n} \psi _{\- ,n-1} (1+\mu
\epsilon )  \right) \right]   \nonumber
\\ & & \ \  +
\left( \prod _{n=1} ^{N} \psi ^{\dagger} _{\+ ,n} \psi _{\+ ,n-1} (1+\mu
\epsilon )  \right)
\left[
\left( \prod _{n=1} ^{N} \psi _{\- ,n} \psi ^{\dagger} _{\- ,n} \right)
+
\left( \prod _{n=1} ^{N} \psi ^{\dagger} _{\- ,n} \psi _{\- ,n-1} (1+\mu
\epsilon )  \right) \right] \Bigg\} \label{jhu}
\end{eqnarray}
In the last step of (\ref{jhu}) we expanded the exponential as in
appendix \ref{Anhang2},
(\ref{B6})--(\ref{B7}).
The last product of  (\ref{jhu})  represents   the double occupation
(see appendix \ref{Anhang2}).
Now we introduce  a prefactor in the measure which
destroys  double occupation and, simultaneously,  allows
the other three states to  survive  with the same weight:
\begin{eqnarray}
Z_{at}&=& \int \left( \prod _{n=1} ^{N}
d\psi _{\+ n}^{\dagger}d\psi _{\+ n}
d\psi _{\- n}^{\dagger}d\psi _{\- n}
\right) \
\prod _{n=1} ^{N}
\left( -\psi _{\+ ,n} ^{\dagger} \psi _{\+ ,n} \psi _{\- ,n} ^{\dagger} \psi
_{\- ,n} -
\psi _{\+ ,n} ^{\dagger} \psi _{\+ ,n} - \psi _{\- ,n} ^{\dagger} \psi _{\- ,n}
\right) \nonumber \\ & & \ \ \ \ \ \ \ \ \times
\exp \left[ -\sum _{\sigma }\sum _{n=1 }^{N} \left( \psi ^{\dagger} _{\sigma
,n}
\psi _{\sigma ,n} -
\psi ^{\dagger} _{\sigma ,n} \psi _{\sigma ,n-1} (1+\mu \epsilon )
\right) \right] \label{fghfgh}
\end{eqnarray}
The quartic term  annihilates all the terms of the exponential
function, except the  zeroth order term.
In order to explore further this  projecting measure
we   re-exponentiate  as
\begin{eqnarray}
&&\left( -\psi _{\+ ,n} ^{\dagger} \psi _{\+ ,n} \psi _{\- ,n} ^{\dagger} \psi
_{\- ,n} -
\psi _{\+ ,n} ^{\dagger} \psi _{\+ ,n} - \psi _{\- ,n} ^{\dagger} \psi _{\- ,n}
\right) \nonumber \\
&&= -\int _{-\pi } ^{\pi } \frac{d \lambda_{n} }{2 \pi }
\exp \left(
i\lambda _{n}\left(
1+ \psi _{\+ ,n} ^{\dagger} \psi _{\+ ,n} \psi _{\- ,n} ^{\dagger} \psi _{\-
,n} -
\psi _{\+ ,n} ^{\dagger} \psi _{\+ ,n} - \psi _{\- ,n} ^{\dagger} \psi _{\- ,n}
\right)
\right) \label{jhjh}
\end{eqnarray}
to find  a contribution to the action:
\begin{eqnarray}
S_{\lambda }&=&  \sum _{n=1} ^{N} i \lambda _{n}
\frac{\epsilon }{\epsilon }
\left(
1+ \psi _{\+ ,n} ^{\dagger} \psi _{\+ ,n} \psi _{\- ,n} ^{\dagger} \psi _{\-
,n} -
\psi _{\+ ,n} ^{\dagger} \psi _{\+ ,n} - \psi _{\- ,n} ^{\dagger} \psi _{\- ,n}
\right) \nonumber \\
&\rightarrow &
\int _{0} ^{\beta } d \tau    i
\lambda (\tau )
\frac{1 }{\epsilon }
\left(
1+ \psi _{\+ ,n} ^{\dagger} \psi _{\+ ,n} \psi _{\- ,n} ^{\dagger} \psi _{\-
,n} -
\psi _{\+ ,n} ^{\dagger} \psi _{\+ ,n} - \psi _{\- ,n} ^{\dagger} \psi _{\- ,n}
\right)
\end{eqnarray}
In the continuum limit $\epsilon \rightarrow 0, $ $ S_{\lambda }$
diverges.
This resembles  an  infinitely strong  local  interaction at each time step.
After translation to operator language and performing
a particle hole transformation (see appendix \ref{Anhang5})  one recognizes
a local interaction term with infinite strength:
\begin{eqnarray}
\frac{\lambda }{\epsilon }(\1 - B_{\+ } B_{\+ }^{\dagger}  - B_{\- } B_{\-
}^{\dagger}+
B_{\+ } B_{\+ }^{\dagger}B_{\- } B_{\- }^{\dagger} )
\rightarrow \frac{\lambda }{\epsilon }C_{\+ }^{\dagger}C_{\+ }C_{\-
}^{\dagger}C_{\- }
\end{eqnarray}

This ad hoc scheme of introducing a measure, which projects into a (fermionic)
subspace, may be embedded in the more general language of section \ref{new}:
The commutation relations of the operators $X^{ab}$ (where $a,b =0,\+,\-$)
clearly constitute an algebra in the constrained space. So we may set up
coherent states as:
\begin{eqnarray}
|G_{0} \rangle &=&\exp ^{\psi _{\+} X^{\+ 0}   +\psi _{\-} X^{\- 0} -h.c.}
|0\rangle \nonumber\\
&=& \left( 1-\frac{1}{2}\sum _{\sigma  } \psi ^{\dagger}
_{\sigma }\psi _{\sigma }
+\frac{3}{4}  \psi ^{\dagger} _{\+ }\psi _{\+ }\psi ^{\dagger}
_{\- }\psi _{\- }  \right) |0\rangle \nonumber\\
&&+ \left( \psi _{\+ }-\frac{1}{2} \psi ^{\dagger}
_{\- }\psi _{\- }\psi _{\+ } \right)|\+\rangle \nonumber\\
&&+ \left(\psi _{\- }-\frac{1}{2} \psi ^{\dagger}
_{\+ }\psi _{\+ }\psi _{\- } \right)|\-\rangle \label{T}
\end{eqnarray}
where $\psi _{\sigma }$ are Grassmann numbers. Now the
projector onto the normalized coherent states reads:
\begin{eqnarray}
P= |G_{0}\rangle \langle G_{0}|&=& X^{00} \left( 1 -
\sum _{\sigma  } \psi ^{\dagger} _{\sigma }\psi _{\sigma } -
2\psi ^{\dagger} _{\+ }\psi _{\+ }\psi ^{\dagger}
_{\- }\psi _{\- }
\right) \nonumber \\
&&+ X^{\+ \+} \left( \psi _{\+ } \psi ^{\dagger} _{\+ }
+ \psi ^{\dagger} _{\+ }\psi _{\+ }\psi ^{\dagger}
_{\- }\psi _{\- } \right) \nonumber \\
&&+ X^{\- \-} \left( \psi _{\- } \psi ^{\dagger} _{\- }
+ \psi ^{\dagger} _{\+ }\psi _{\+ }\psi ^{\dagger}
_{\- }\psi _{\- } \right)
\end{eqnarray}
For the resolution of unity we have to find a measure $\mu
(\psi ^{\dagger} _{\+ }, \psi _{\+ }, \psi ^{\dagger}
_{\- }, \psi _{\- } )$ which satisfies:
\begin{eqnarray}
\int d\psi _{\+ }d\psi ^{\dagger} _{\+ }d\psi _{\- }d\psi ^{\dagger}
_{\- }   \
\mu (\psi ^{\dagger} _{\+ }, \psi _{\+ }, \psi ^{\dagger}
_{\- }, \psi _{\- } ) P =\1\label{star}
\end{eqnarray}
Obviously, any measure consisting of a single product of the Grassmann numbers
will be inadequate to solve (\ref{star}).
It turns out that $\mu $ has to be of an additive form
\begin{eqnarray}
\mu
(\psi ^{\dagger} _{\+ }, \psi _{\+ }, \psi ^{\dagger}
_{\- }, \psi _{\- } )= - \left( \psi ^{\dagger} _{\+ }\psi _{\+ }\psi
^{\dagger}
_{\- }\psi _{\- }+
\psi ^{\dagger} _{\+ }\psi _{\+ }+\psi ^{\dagger}
_{\- }\psi _{\- }
\right) \label{starstar}
\end{eqnarray}
in order to ensure the resolution of the unity operator.

The Berry phase has to be calculated in the discrete version.
Terms in fourth order of the Grassmann fields arise, but are eliminated by
multiplication with $\mu $ of (\ref{starstar}).
The surviving contribution has the canonical form
$\sum _{\sigma } \psi ^{\dagger }_{\sigma } \partial _{\tau }\psi _{\sigma }$,
as in (\ref{fghfgh}).
Having fixed the measure and the Berry phase, the functional integral
formalism is established.
One might worry that the derivation of the Hamiltonian terms of the action
always
involves higher orders in the field variables; e.g.\
the kinetic term as derived with (\ref{T}) is
\begin{eqnarray}
& & \sum _{<i,j>} t_{ij}
\left( \psi ^{\dagger} _{\sigma ,i, n }
- \frac{1}{2} \psi ^{\dagger} _{\sigma ,i, n }
\psi ^{\dagger} _{-\sigma ,i, n-1 }\psi ^{\dagger} _{-\sigma ,i, n-1 }
- \frac{1}{2} \psi ^{\dagger} _{\sigma ,i, n }
\psi ^{\dagger} _{-\sigma ,i, n }\psi ^{\dagger} _{-\sigma ,i, n } \right)
\nonumber \\ & &   \ \   \ \ \   \   \
\left( \psi ^{\dagger} _{\sigma ,j, n }
- \frac{1}{2} \psi ^{\dagger} _{\sigma ,j, n }
\psi ^{\dagger} _{-\sigma ,j, n-1 }\psi ^{\dagger} _{-\sigma ,j, n-1 }
- \frac{1}{2} \psi ^{\dagger} _{\sigma ,j, n }
\psi ^{\dagger} _{-\sigma ,j, n }\psi ^{\dagger} _{-\sigma ,j, n } \right)
\end{eqnarray}
Fortunately,  the measure $\mu $ eliminates all fourth and higher order terms,
similarly to the calculation of
the Berry phase.

Before we try  to derive an effective Lagrangian for large $U$, we
should summarize the projection techniques of the last two sections:
It turned out that it is possible to construct normalized coherent states which
lie in the constrained space and resolve the unity.
But their measure is of an additive form, i.e., a polynomial of
the Grassmann variables for up and down spin.
The minimal number of complex Grassmann fields for such a resolution in the
constrained space
is two, so that the proposed coherent state  does not render the path integrals
more
tractable as compared to the canonical formalism with two Grassmann fields
in the full space:
The local interaction is suppressed but at the expense of a most
inconvenient measure.
Therefore,
coherent states should be set up in the full space in which we may reduce the
number of Grassmann fields to one.
Additional complex fields have to be introduced for spin and pseudospin.
Since the local interaction denotes a `magnetic field'
in  the  pseudospin space, we investigated an elaborate saddle point technique
for spins in magnetic fields.
It serves to constrain the excitations   to the considered subspace
by sending the magnetic field   to infinity in
a controlled way:
The pseudospin fields may be fixed to their `subspace  direction'
everywhere except in the Berry phase where small fluctuations about
this direction have to be kept, even for infinite $U$.

\section{Cumulant Expansion for Large  Local Interaction}

Now we will show how to gain the effective action to lowest order in $t/U$
using the coherent states (\ref{ioioio}) and the projection method of section
\ref{pro}. The effective action will be similar, but not identical, to the
action
derived by Schulz through a Hubbard Stratonovich decoupling scheme.

The Lagrangian for the Hubbard model was introduced in section \ref{app4}:
\begin{eqnarray}
L &=&\sum _{i}  \ \Big(
-\mu +i\dot{\varphi _{i}}\sin^{2}{\vartheta _{i}}  +\nonumber \\
& & +   \left.
\psi^{\dagger}_{i} \left( \partial _{\tau } + \frac{U}{2}
+ i \dot{\phi_{i}}\sin^{2} {\theta _{i}}
-i \dot{\varphi_{i}}\sin^{2} {\vartheta _{i}}  +
\left( \mu -\frac{U}{2} \right)
\cos (2\theta _{i})
\right)
\psi _{i}
\right)   + L_{t} \label{gerger}
\end{eqnarray}
Applying the projection prescription (\ref{prescription}), the Lagrangian
reads for  $U=\infty $:
\begin{eqnarray}
\overline{L} =\sum _{i} \ \left(
-\mu +i\dot{\varphi _{i}}\sin^{2}{\vartheta _{i}} +
\psi^{\dagger} _{i}\left( \partial _{\tau } + \mu + i \dot{\phi_{i}}\sin^{2}
{\theta _{i}}
\Big| _{\mbox{\scriptsize north pole}}
-i \dot{\varphi_{i} }\sin^{2} {\vartheta _{i} }
\right)
\psi _{i}
\right)    + \overline{L} _{t}   \label{gerger2}
\end{eqnarray}
where $\overline{L}_{t}$ restricts hopping to the exchange
of a hole with either a $\+ $-spin or a $\- $-spin on the
nearest neighbor site:
\begin{eqnarray}
\overline{L} _{t}= \sum _{<i,j>}
t_{ij}\psi_{i} \psi_{j}^{\dagger}
\left[ \left(\sin{\vartheta_{i}}\sin{\vartheta_{j}}
e^{-i\left( \varphi _{i}-\varphi _{j}\right)}
+ \cos{\vartheta_{i}} \cos{\vartheta_{j}}\right)\right]
\end{eqnarray}\\

The path integral with  $\overline{L} $ from (\ref{gerger2}) may be
cast  into a more conventional form without a prescription.
Through a redefinition of the measure
\begin{eqnarray}
&&{\cal D } \left[ \theta ( \tau ),
\phi (\tau ), \vartheta ( \tau ),
\varphi (\tau ), \psi ^{\dagger}(\tau),\psi (\tau) \right]
\nonumber \\ &&
\rightarrow
{\cal D } \left[ \theta ( \tau ),
\phi (\tau ), \vartheta ( \tau ),
\varphi (\tau ), \psi ^{\dagger}(\tau),\psi (\tau) \right]
\exp{\left[ -\sum _{i}\int _{0}^{\beta }d \tau \ \left( \psi ^{\dagger}
_{i}\psi_{i}
\ i\  \dot{\phi_{i}}\sin^{2} {\theta _{i}}
\Big| _{\mbox{\scriptsize north pole}}  \right) \right] }
\end{eqnarray}
the projected Lagrangian  acquires the form:
\begin{eqnarray}
L_{U=\infty } &=& \sum _{i}  \ \left(
-\mu +i\dot{\varphi _{i}}\sin^{2}{\vartheta _{i}} +
\psi^{\dagger}_{i} \left( \partial _{\tau } + \mu
-i \dot{\varphi_{i}}\sin^{2} {\vartheta _{i}}
\right)
\psi   _{i}
\right)  \nonumber \\ & & +
\sum _{<i,j>}  \  t_{ij}\psi_{i} \psi_{j}^{\dagger}
\left[ \left(\sin{\vartheta_{i}}\sin{\vartheta_{j}}
e^{-i\left( \varphi _{i}-\varphi _{j}\right)}
+ \cos{\vartheta_{i}} \cos{\vartheta_{j}}\right)\right]
\end{eqnarray}
This form of the path integral would be more useful
if we were able to integrate over the purely kinematical fields $\theta $ and
$\phi$ first. \\

In the next stage,  we will explicitly calculate local propagators
in order to gain an effective Lagrangian
which is generated by a  hopping expansion.
The dynamical processes of this  effective Lagrangian   lie
in the constrained space only.
We will find the propagator  by integrating
first over the Grassmann fields and then the (pseudo)  spin fields.
As an introductory exercise we calculate $Z$ for the atomic limit with
$\overline{L} _{0}= \overline{L} - \overline{L} _{t}$
(analogously to section \ref{AusintatLim}):
\begin{eqnarray}
Z&=& \int {\cal D} \left[  \vartheta (\tau ),\varphi (\tau )
,\theta (\tau ), \phi (\tau ),\psi (\tau ),\psi ^{\dagger}(\tau )\right]
\exp \left[ -\int _{0}^{\beta }\overline{L} _{0} \ d\tau  \right]
\nonumber \\
&=&\int \left(\prod _{l=1}^{N} \frac{1}{\pi ^{2} }d \vartheta _{l} d
\varphi _{l}
  \right) \
\prod _{l=1}^{N} \langle {\bf n}_{l} | {\bf n}_{l-1}  \rangle e^{\mu \beta }
+\nonumber \\
& & +
\int \left( \prod _{l=1}^{N} \frac{1}{\pi ^{2} } d \theta  _{l}
d \phi _{l}   \right) \,
e^{\mu \beta }
\prod _{l=1}^{N}
\langle {\bf N}_{l}  | \left(
\1
+\mu \sigma _{z} \epsilon \right) P_{\mbox{\scriptsize north pole}}
|{\bf N}_{l-1}  \rangle  \nonumber \\
&=& 2e^{\mu \beta }+1
\end{eqnarray}
where $P_{\mbox{\scriptsize north pole}} =\left(
\begin{array}{cc} 0 &0  \\  0 & 1
\end{array} \right) $ and we have used the fact that
only the two complete Grassmann chains survive. \\

Similarly,  we may evaluate the local propagators necessary for a cumulant
resummation.
In order to check the consistency of the calculation we compare the propagator
obtained with $L$ to  the projected propagator assigned to   $\overline{L}$. \\

First we    expand the functional in terms of the hopping
parameter $t$.
Odd orders in the hopping expansion vanish because  they
contain  an odd number of Grassmann numbers at each of the sites involved
in  the hopping process.
In  second order we  average with the original action $S_{at}=
\int _{0} ^{\beta }  d \tau \ (L-L_{t})$.
For example the propagator $Z  \cdot \langle \mbox{ T} \ X_{i=0}^{\- 0 }(\tau
_{m-1})
X _{i=0}^{0 \-}(\tau _{n})\rangle $ is:
\begin{eqnarray}
&&\int {\cal D} \left[
\vartheta  (\tau ),\varphi  (\tau ),\theta  (\tau ),\phi  (\tau ),
\psi  ^{\dagger} (\tau ), \psi  (\tau )\right]\nonumber \\
&&  \ \ \ \ \ \ \  \ \ \ \
\times \psi _{m-1} \psi  ^{\dagger}_{n}
\cos (\theta _{m-1})  \cos (\theta _{n}) \cos (\vartheta _{m})
\cos (\vartheta _{n-1})
\exp \left[ -S_{at} \right] \label{r83}
\end{eqnarray}
where $n \le m $ is assumed.
With the identity
\begin{eqnarray}
\cos (\theta _{n})  \cos (\theta _{m-1}) =
\langle {\bf N}_{m-1} |
\left(
\begin{array}{cc}
0&0 \\ 0&1
\end{array}
\right)
|{\bf N}_{n} \rangle
\end{eqnarray}
we get
\begin{eqnarray}
& &e^{\mu \beta }\int {\cal D} \left[
\vartheta  (\tau ),\varphi  (\tau ),\theta  (\tau ),\phi  (\tau ),
\psi  ^{\dagger} (\tau ), \psi  (\tau )\right]
\ \psi _{m-1} \psi  ^{\dagger}_{n} \nonumber \\
& & \ \  \  \times
\left(  \prod _{l = m} ^{n-1}    \psi _{l} \psi  ^{\dagger}_{l} \right)
\left( \prod _{l = m+1} ^{n-1}   \langle {\bf
n}_{l}   |{\bf n}_{l-1}
\rangle
\right)
\langle {\bf n}_{m}  |
\left(
\begin{array}{cc}
0&0 \\ 0&1
\end{array}
\right)
|{\bf n}_{n-1} \rangle \nonumber \\
& & \ \ \ \times \prod _{l = n+1} ^{m-1} \left( \psi  ^{\dagger}_{l} \psi
_{l-1}
\langle{\bf N}_{l}  | \1 - \left(\frac{U}{2} \1
-(\mu - \frac{U}{2})\sigma  _{z} \right) \epsilon    |{\bf N}_{l-1}  \rangle
\right)
\langle {\bf N}_{n}  |
\left(
\begin{array}{cc}
0&0 \\ 0&1
\end{array}
\right)
|{\bf N}_{m-1} \rangle  \nonumber \\
&=&
e^{\mu \beta }\int {\cal D} \left[
\psi  ^{\dagger} (\tau ), \psi  (\tau )\right]
\ \psi _{m-1} \psi  ^{\dagger}_{n}
\Sp \left[
\left(
\begin{array}{cc}
0&0 \\ 0&1
\end{array}
\right)
\right]
\prod _{l = m} ^{n-1}  \psi _{l} \psi  ^{\dagger}_{l}
\prod _{l = n+1} ^{m-1} \psi  ^{\dagger}_{l} \psi _{l-1}   \nonumber \\
& & \times \Sp \left[
\left(
\begin{array}{cc}
0&0 \\ 0&1
\end{array}
\right)
\exp \left[
\left( -\frac{U}{2} \1
+(\mu - \frac{U}{2}) \sigma  _{z} \right) \epsilon (m-n-1)
\right]  \right] \nonumber \\
& = &
e^{\mu \beta }e^{-\mu (m-n-1) \epsilon } \label{r85}
\end{eqnarray}
where the following convention  is introduced:
if there is a sum (product), whose upper index is smaller than
its lower index,  the sum (product)  denotes  a sum (product)
from the lower index to $\beta $ and from 0 to the upper index.
With this  convention  we can also prove the result to be true in the case
of $n\ge m$.

By exploiting  the identity
\begin{eqnarray}
\sin (\theta _{m-1})  \sin (\theta _{n}) e^{i( \phi _{m-1} - \phi _{n})}=
\langle {\bf N}_{n}  |
\left(
\begin{array}{cc}
1&0 \\ 0&0
\end{array}
\right)
|{\bf N}_{m-1} \rangle
\end{eqnarray}
we are able to calculate remaining term from the second order
cumulant expansion. It  which will generate virtual double occupations
in the effective action below.
\begin{eqnarray}
& &\int {\cal D} \left[
\vartheta  (\tau ),\varphi  (\tau ),\theta  (\tau ),\phi  (\tau ),
\psi  ^{\dagger} (\tau ), \psi  (\tau )\right]
\ \psi _{m-1} \psi  ^{\dagger}_{n} \nonumber \\
& & \ \ \ \ \times
\sin (\theta  _{m-1})  \sin (\theta _{n}) e^{ i(\phi _{m-1} -\phi
_{n} ) }
\cos (\vartheta _{m}) \cos (\vartheta _{n-1})
e^{-S_{at}}\nonumber \\
& &=
e^{\mu \beta }\int {\cal D} \left[
\psi  ^{\dagger} (\tau ), \psi  (\tau )\right]
\ \psi _{m-1} \psi  ^{\dagger}_{n} \
\Sp \left[
\left(
\begin{array}{cc}
0&0 \\ 0&1
\end{array}
\right)
\right]
\prod _{l = m} ^{n-1}  \psi _{l} \psi  ^{\dagger}_{l}
\prod _{l = n+1} ^{m-1} \psi  ^{\dagger}_{l} \psi _{l-1} \nonumber \\
& & \ \ \ \  \times \Sp \left[
\left(
\begin{array}{cc}
1&0 \\ 0&0
\end{array}
\right)
\exp \left[
\left(- \frac{U}{2} \1
+(\mu - \frac{U}{2})\sigma  _{z} \right) \epsilon (m-n-1)  \right] \right]
\nonumber \\
& & = e^{\mu \beta }e^{(\mu -U ) (m-n-1) \epsilon }
\end{eqnarray}
We have to pay attention to the correct sequence
of   the limits, i.\ e.\  to take first $U\rightarrow
\infty $ and then $\epsilon \rightarrow 0$ (for $\mu \ll U$).
In the case of large $U$, it is an excellent approximation   to
substitute for the last result a Kronecker delta with weight
given by the integral of the exponential over all positive
times $\tau =(m-n-1)\epsilon $ \cite{Schulz}:
\begin{eqnarray}
\approx \frac{1}{U} e^{\mu \beta } \delta _{m-1,n}
\end{eqnarray}

\vspace{.7cm}
Now we  calculate the analogous  averages using the projected action,
$\overline{S} _{at}=\int _{0} ^{\beta }
d \tau \ (\overline{L}-\overline{L}_{t})$. The calculation corresponds to
the preceding:
\begin{eqnarray}
& & \int {\cal D} \left[
\vartheta  (\tau ),\varphi  (\tau ),
\psi  ^{\dagger} (\tau ), \psi  (\tau )\right]
\ \psi _{m-1} \psi  ^{\dagger}_{n}
\cos (\vartheta _{m}) \cos (\vartheta _{n-1})
e^{-\overline{S} _{at}} \nonumber \\
& & = \int {\cal D} \left[
\vartheta  (\tau ),\varphi  (\tau ),
\psi  ^{\dagger} (\tau ), \psi  (\tau )\right]
\ \psi _{m-1} \psi  ^{\dagger}_{n} e^{\mu \beta }
 \langle {\bf n}_{m}  |
\left(
\begin{array}{cc}
1&0 \\ 0&0
\end{array}
\right)
|{\bf n}_{n-1} \rangle
\nonumber \\
& & \ \ \ \
\prod _{l = m} ^{n-1} \left( \psi _{l} \psi  ^{\dagger}_{l}\right)
\prod _{l = m+1} ^{n-1} \left(
\langle {\bf n} _{l}| {\bf n}  _{l-1}\rangle \right)
\prod _{l = n+1} ^{m-1} \psi  ^{\dagger}_{l} \psi _{l-1}
\langle {\bf N} _{l}| (\1 - \mu \epsilon )P_{\mbox{\scriptsize north
pole}}|{\bf N}  _{l-1}\rangle
\nonumber \\
& &=e^{\mu \beta }e^{-\mu (m-n-1) \epsilon }
\end{eqnarray}
which is obviously equivalent to (\ref{r83})--(\ref{r85}).
Therefore we are led to introduce the following translation
rules regarding the transition  from the unconstrained to the constrained
(infinite $U$) problem.
\begin{eqnarray}
&\langle&\psi ^{\dagger}_{n} \psi _{m-1}\cos (\theta _{n}) \cos (\theta _{m-1})
... \rangle
_{S_{at}}
\rightarrow \langle  \psi ^{\dagger}_{n} \psi _{m-1}...  \rangle_{\overline{S}
_{at}}\label{Regeln111}\\
&\langle&\psi ^{\dagger}_{n} \psi_{m-1} \sin (\theta _{n}) \sin  (\theta
_{m-1}) ... \rangle
_{S_{at}}
\rightarrow \frac{1}{U} \delta _{n,m-1} \langle ...  \rangle_{\overline{S}
_{at}} \label{Regeln222}
\end{eqnarray}
The ellipses stand for the various spin variables.
(\ref{Regeln111}) and (\ref{Regeln222}) exemplify
how to gain all the required propagators.
The hopping expansion has to be done to fourth order. After
using these translation rules  we re-exponentiated  to
obtain the following effective Lagrangian:
\begin{eqnarray}
L_{eff}&=&\sum _{i} L_{0}
\nonumber \\ & &-t
\sum _{<i,j>} \psi_{i,n-1} \psi_{j,n}^{\dagger}
\alpha ({\bf n} _{i},{\bf n} _{j} )
\nonumber \\ & & + \frac{t^{2} }{U}
\sum _{<i,j>}\left(
\psi_{i,n-1} \psi_{i,n}^{\dagger}
+\psi_{j,n-1} \psi_{j,n}^{\dagger}
\right)
\left(
1-{\bf n} _{i} {\bf n} _{j}
\right) \nonumber \\ & &
+\frac{t^{2} }{U}
\sum _{<i,j,k>}
\psi_{i,n-1} \psi_{k,n}^{\dagger}
\alpha ({\bf n} _{i},{\bf n} _{j} )
\alpha ({\bf n} _{j},{\bf n} _{k} )
\end{eqnarray}
Here $L_{0}=\overline{L}-\overline{L}_{t}$ and
\begin{eqnarray}
\alpha ({\bf n} _{i},{\bf n} _{j} )=
\sqrt{\frac{1}{2}   \left( 1+{\bf n} _{i}{\bf n} _{j}\right) } \ \
\exp{\left[ i {\cal A } ({\bf n} _{i},{\bf n} _{j},\hat{\bf z }  ) /2
\right] }
\end{eqnarray}
${\cal A } ({\bf n} _{i},{\bf n} _{j},\hat{\bf z }  )$
is the area of the spherical triangle defined by the three unit vectors,
and $\hat{\bf z }$ points to the north pole of the unit sphere.
$<i,j,k>$ denotes the following convention: $j$ has to be a nearest neighbor
of  $i$, and $k$ a nearest neighbor of $j$, excluding $k=i$.
To compare this  effective Lagrangian with   Schulz
\cite{Schulz}, we have to perform a particle-hole transformation.
On the operator  level   the well known anticommutator relation
\begin{eqnarray}
C^{\dagger} C = \1 - C C^{\dagger}
\end{eqnarray}
holds. In the path integral formalism we work with the corresponding
`anticommutator relation':
\begin{eqnarray}
& &\int \left( \prod _{n=1} ^{N} d \psi _{l} d \psi ^{\dagger}_{l}\right) \
\psi ^{\dagger} _{n}\psi _{n-1}\exp \left[ -\sum  _{l=1} ^{N}
\psi ^{\dagger} _{l}(\psi  _{l}-\psi  _{l-1}(1+\mu \epsilon ) )
\right] \nonumber \\ & & =
\int \left( \prod _{n=1} ^{N} d \psi _{l} d \psi ^{\dagger}_{l}\right) \
\left( 1 - \psi  _{n}\psi _{n} ^{\dagger} \right) \exp \left[ -\sum  _{l=1}
^{N}
\psi ^{\dagger} _{l}(\psi  _{l}-\psi  _{l-1}(1+\mu \epsilon ) )
\right] \label{rrr}
\end{eqnarray}
(\ref{rrr}) is proved by using relations of appendix \ref{Anhang5}.
Obviously, the  information about the anticommutator is
contained in the
time index (see also \cite{Schulmann}).
The transformation results in:
\begin{eqnarray}
{\cal L}={\cal L}_{0}+{\cal L}_{t}+{\cal L}_{J}+{\cal L}_{pair}
\end{eqnarray}
where
\begin{eqnarray}
{\cal L}_{0} &=&\frac{1}{\epsilon}\sum _{i} \left(
-\mu \epsilon + 1   -
\langle {\bf n }_{i,n}  |  {\bf n }_{i,n-1 }\rangle
+\psi ^{\dagger} _{i,n} \psi _{i,n}
\right. \nonumber \\
& &\left. +\psi ^{\dagger} _{i,n} \psi _{i,n-1}\left(
\mu \epsilon - 1 +
\langle {\bf n }_{i,n}  |  {\bf n }_{i,n-1 }\rangle -
\langle {\bf N }_{i,n}  |  {\bf N }_{i,n-1 } \rangle
\Big| _{\mbox{\scriptsize north pole}}
\right)
\right)
\end{eqnarray}
contains the Berry phase which differs from \cite{Schulz}
by pseudo spin fluctuations about  the north pole,
\begin{eqnarray}
{\cal L}_{t}=-t
\sum _{<i,j>} \psi ^{\dagger}_{i,n} \psi_{j,n-1}
\alpha ({\bf n} _{i},{\bf n} _{j} )
\end{eqnarray}
is the hopping term,
\begin{eqnarray}
{\cal L}_{J}= \frac{t^{2}}{U}
\sum _{<i,j>}\left(2-
\psi_{i,n}^{\dagger} \psi_{i,n}
-\psi_{j,n}^{\dagger} \psi_{j,n}
\right)
\left(
1-{\bf n} _{i} {\bf n} _{j}
\right)
\end{eqnarray}
denotes the Heisenberg term, and
\begin{eqnarray}
{\cal L}_{pair}=
\frac{t^{2}}{U}
\sum _{<i,j,k>}
\psi_{i,n}^{\dagger} \psi_{k,n}
\alpha ({\bf n} _{i},{\bf n} _{j} )
\alpha ({\bf n} _{j},{\bf n} _{k} )
\end{eqnarray}
is the pair hopping term which represents hopping to the next nearest
neighbor site and
thereby transports a pair of spins.
We refer  the reader to the literature for a discussion of these terms, e.g.\
\cite{Fradkin,Schulz}.

Schulz   derived an additional term:
 \begin{eqnarray}
\frac{1}{4U}\sum _{j} (1-\psi ^{\dagger}_{j} \psi _{j} ) \dot{ {\bf n }}
^{2}_{j}
\end{eqnarray}
Such a term is not generated in our formalism because
no operation will produce  a quadratic
time derivative.

\section{Conclusions}

In this paper we addressed the question of how to set up a path integral
formalism
in a
constrained Fock space.
We only tackled methods which do not involve additional `graded constraints'
as in the slave boson method.
The projection was implemented through either, (i) a suitable modification of
the measure of the path integral, or, (ii) through adequately generalized
coherent states.

The first approach allows the use  of Grassmann fields which correspond
to the canonical
creation and annihilation operators of electrons.
A `projecting measure' was introduced which explicitly projects at every
time step.

Alternatively, one might want to consider generalized
coherent states which lie in the constrained space, exclusively.
Such normalized states were constructed explicitly
but they resolve the unity operator only if the number
of complex Grassmann fields is equal to the number of spin states.
But even then, the path integration is awkward because
the Grassmann valued measure is of an additive form which turns
out to be exactly the `projecting' measure of the first approach.
This implies that  normalized coherent states should
be set up in the full Fock space first. We presented such states, involving
only one
(complex) Grassmann field  and the smallest number of complex fields
possible, i.\ e.\ one for the spin and
one for the charge degree  of freedom.
The `charge field', which here signifies transitions between
states with empty and doubly occupied sites, was shown to be a pseudo spin
coupled to a `magnetic field',
linear in the local interaction parameter.
Accordingly, this approach had to establish a procedure how to send
the pseudo  magnetic field to infinity in order to
implement the constraint of no double occupancy.
This procedure fixes the pseudo spin field at the north pole
in the Hamiltonian terms of the action but keeps fluctuations
around the north pole in the  Berry phase.
It seems to be a generic feature of such constrained systems
that forces us to keep the Berry phase fluctuations in the unconstrained space,
in order to work with
`normalizable' path integrals.
Finally, this method was elaborated and tested on the Hubbard model to derive
an effective Lagrangian for strong local interaction. \\

We would like to thank
P.~W\"olfle for valuable comments and his steady encouragement.
We are grateful to N.~Andrei, T.A.~Costi, F.R. Klinkhamer,  A.~Rosch, and
A.E.~Ruckenstein for   stimulating discussions.
Part of this work was supported by the Deutsche Forschungsgemeinschaft, SFB
195.

\appendix

\section{Spin path integral }
\label{Anh1}

A spin coherent state  is given by\cite{Perelomov}:
\begin{eqnarray}
| {\bf n } \rangle =\left(
\begin{array}{ll}
\sin \left( \vartheta \right) e^{i\varphi }  \\
\cos \left( \vartheta  \right)
\end{array}
\right)
\end {eqnarray}
Furthermore  we can find a resolution of unity and a trace
with this  overcomplete set of coherent states:
\begin {eqnarray}
\1 &=& \int d \mu \left(  {\bf n }\right) \ | {\bf n }\rangle \langle
{\bf n } | \nonumber \\
&=&
\frac{1}{\pi ^{2}} \int _{0} ^{\pi} d \vartheta  \int _{0} ^{2\pi} d \varphi
\left(
\begin{array}{ll}
\sin ^{2} (\vartheta ) & \sin (\vartheta ) \cos (\vartheta )e^{i\varphi } \\
\sin (\vartheta ) \cos (\vartheta ) e^{-i\varphi } & \cos ^{2} (\vartheta)
\end{array}
\right)
=\left(
\begin{array}{ll}
1 & 0 \\ 0 & 1
\end{array}
\right) \nonumber \\ & & \nonumber \\
\Sp \ \left( A \right)  &= &
\frac{1}{ \pi ^{2}} \int _{0} ^{\pi} d \vartheta  \int _{0} ^{2\pi} d \varphi
\  \langle {\bf n } | A | {\bf n }\rangle
\end {eqnarray}
Now    a path integral is derived using the Trotter formula and
introducing
an imaginary time index:
\begin{eqnarray}
Z&=&\Sp \ e^{-\beta H} \nonumber \\
&=&
\lim _{N \rightarrow \infty }\Sp \left(  e^{-\epsilon H } \1 _{1}
e^{-\epsilon H } \1 _{2}
e^{-\epsilon H } ..... \1 _{N-3 }
e^{-\epsilon H }  \1 _{N-2}e^{-\epsilon H } \1 _{N-1} e^{-\epsilon H } \right)
\nonumber \\  &=&
\lim _{N \rightarrow \infty } \int \prod _{l=1 } ^{N} d \mu \left(
{\bf n }_{l} \right) \
\prod _{l=1 } ^{N}
\langle {\bf n } _{l}| \1 -\epsilon H |{\bf n }_{l-1} \rangle
\nonumber \\
&=&
\lim _{N \rightarrow \infty } \int \prod _{l=1 } ^{N} d \mu \left(
{\bf n }_{l} \right) \
\exp \left[ { - \sum _{l=1 } ^{N}\left(
 \epsilon
\langle {\bf n } _{l} |\partial _{\tau }|{\bf n }_{l} \rangle
+\epsilon \langle {\bf n } _{l} |H|{\bf n }_{l} \rangle  +
{\cal O } \left( \epsilon ^{2 } \right)
\right) } \right]
\end{eqnarray}
In the last step    the continuum limit was taken.
Thereby the discrete time index ${\bf n }_{l}$  became a continuum variable
${\bf n }(\tau ) $ with $\tau = l \epsilon $ and the measure  is
$d\mu ({\bf n }_{l}) = \frac{d \vartheta _{l} d \varphi _{l } }{\pi ^{2}} $.
The
continuum path integral is  written as:
\begin{eqnarray}
Z = \int {\cal D }
\left[ \vartheta \left( \tau \right), \varphi \left( \tau \right)\right]
\exp \left[ -\int _{0} ^{\beta }\left( i \dot{\varphi } \sin ^{2} \left(
\vartheta \right) +
\langle {\bf n }| H |{\bf n } \rangle
\right) d \tau \right]
\end{eqnarray}
We finish this appendix with a note of warning:
Usually   a   path integral with  complex fields
is evaluated through a  saddle point approximation.
In the spin case it will  break down.
Consider the spin in a magnetic field proportional to $B$.
The Lagrangian is:
\begin{eqnarray}
L= i \dot{\varphi } \sin ^{2} \left(
\vartheta \right) + B \cos (2\vartheta ) \label{SpinimMagnetfeld}
\end{eqnarray}
The general solution of the classical equations of motion
restricts the spin dynamics to north and south pole if
the periodic boundary conditions are taken care of.
The classical action   is $S_{cl}= \pm  B\beta $, for   north and south  pole,
respectively.
Although the saddle point solution    yields
the  correct result,  fluctuations
around the    classical path   are divergent, already  in  lowest order.
This originates  from  an expansion of the action
about its maximum at the south pole.
A path on the sphere,   on which the spins live, can  be described only
by two complex numbers at each space and time step.
In  section \ref{pro}
methods  were devised  in order to generalize the conventional
saddle point approximation for spin path integrals.

\section{Fermionic Path Integral}
\label{Anhang2}

In this appendix we want to derive a path integral
for a system with spinless fermions using
normalized coherent states.
We take the following definition for coherent states:
\begin {eqnarray}
|G \rangle =e^{\psi C^{\dagger} -C \psi ^{\dagger} }
| 0 \rangle   \label{tyty}
\end {eqnarray}
Because of the anticommuting properties of the Grassmann numbers
the exponential function is trivially
determined to be:
\begin {eqnarray}
| G \rangle =\left( 1 +  \frac{1}{2}  \psi \psi ^{\dagger} \right) | 0 \rangle
+ \psi | 1 \rangle \label{aghi0}
\end {eqnarray}
This state is normalized
\begin {eqnarray}
\langle G|G \rangle = \left( \langle 0 | \left( 1 +  \frac{1}{2}
                                             \psi \psi ^{\dagger} \right)
                   +\langle 1  | \, \psi ^{\dagger}
                     \right)
\left( \left( 1 +  \frac{1}{2}  \psi \psi ^{\dagger} \right) | 0 \rangle
 + \psi |  1  \rangle \right) = 1
\end {eqnarray}
and we can easily find a resolution of the unity operator
\begin {eqnarray}
\int d \psi ^{\dagger} d \psi |G \rangle \langle G|
&=&
\int d \psi ^{\dagger} d \psi \bigg\{
\left( 1+ \psi \psi ^{\dagger} \right) | 0 \rangle      \langle 0 |
+ \ \psi \psi ^{\dagger} | 1 \rangle \langle  1 |   +  \psi  | 1  \rangle
\langle 0|
\ - \psi ^{\dagger}  | 0 \rangle \langle  1 | \bigg\} \nonumber\\
&=&  | 0 \rangle \langle 0| \ + \  | 1  \rangle \langle  1 | =  \1
\label{aghi1}
\end {eqnarray}
and  the  trace is:
\begin{eqnarray}
\Sp \ A = \int d \psi ^{\dagger} d \psi
\left(
\langle 0 | \left( 1 +  \frac{1}{2}  \psi \psi ^{\dagger} \right)
-\langle  1 | \psi  ^{\dagger}
\right)
A
\left(
\left( 1 +  \frac{1}{2}  \psi \psi ^{\dagger} \right) | 0 \rangle   + \psi | 1
\rangle
\right) \label{Spurvz} \label{aghi2}
\end{eqnarray}
In the same manner as in   appendix \ref{Anh1}
we   derive a path integral using the Trotter formula\footnote{Heed the
footnote of section
\ref{AusintatLim}}:
\begin{eqnarray}
Z=\int  \left( \prod _{n=1} ^{N } d \psi ^{\dagger}_{n} d \psi_{n}\right)
\exp \left[ -\sum _{n=1} ^{N }  \epsilon
\left( \psi ^{\dagger} _{n} \left( \psi _{n } - \psi _{n-1} \right) +
\langle G_{n} | H |G_{n-1} \rangle  \right)
\right] \label{B6}
\end{eqnarray}
If we only consider a chemical potential $\mu C^{\dagger}C$, we have to replace
$\langle G_{n} | H |G_{n-1} \rangle \rightarrow -\mu \psi ^{\dagger}_{n}\psi
_{n-1}$.
The evaluation of  the path integral  is
equivalent to  the  calculation of a determinant.
Alternatively, we can expand the exponential function.
Then we  have to collect  a large  number of terms which seem to be difficult
to handle.
But, only the complete chains  of Grassmann numbers will
survive the Grassmann integrations.
Thereby `complete chain' denotes   a product of Grassmann numbers over
all time steps $ \prod _{n=1} ^{N }\psi _{n}
\psi ^{\dagger} _{n} $.
This implies that  the partition function may be written   in the
following way:
\begin{eqnarray}
Z&=& \int \left( \prod _{n=1} ^{N }  d \psi ^{\dagger}_{n} d \psi_{n}\right)
\left(
\prod _{n=1} ^{N }\psi _{n}
\psi ^{\dagger} _{n}
-\prod _{n=1} ^{N }
\psi ^{\dagger} _{n}\psi _{n-1}\left( 1+\mu \epsilon \right)
\right)\nonumber \\
&=& 1+\prod _{n=1} ^{N }\left( 1+\mu \epsilon \right) = 1+  e^{\mu \beta }
\label{B7}
\end{eqnarray}
The assumption of the continuum limit to   exist
requires  $\mu (\tau )$ to be a smooth function.
With this assumption we are able to calculate the
partition function of a time dependent (chemical) potential:
\begin{eqnarray}
Z= 1+\prod _{n=1} ^{N }\left( 1+\mu_{n} \epsilon \right) = 1+  e^{\int_{0} ^{
\beta } \mu \left( \tau \right) d \tau
} \label{zeitabhchemPot}
\end{eqnarray} \\

In order to set up more involved path integrals
we would like to investigate the question
of the minimal number of complex fields.
Specifically, one might have the idea to introduce an additional
complex field in (\ref{tyty}).
\begin {eqnarray}
|G\rangle  = e^{\alpha \psi C^{\dagger} -\alpha ^{\dagger} C \psi ^{\dagger} }
| 0 \rangle = \left( 1 +  \frac{1}{2}  \alpha \alpha ^{\dagger}
\psi \psi ^{\dagger} \right) | 0 \rangle  + \alpha \psi | 1  \rangle
\end {eqnarray}
The state is also  normalized   and we can resolve  the unity:
\begin{eqnarray}
\1 =\int d \psi ^{\dagger} d \psi \int d \alpha  d \alpha ^{\dagger} \mu
\left( \alpha \right) | G\rangle \langle G|
\end{eqnarray}
Here, the measure has to  obey the relation:
\begin{eqnarray}
\int d \alpha  d \alpha ^{\dagger} \ \mu \left( \alpha \right)  \
\alpha ^{\dagger} \alpha = 1
\end{eqnarray}
The following  Berry phase arises in the path integral:
\begin{eqnarray}
{\cal L} _{o}= \alpha ^{\dagger} \psi ^{\dagger}  \partial _{\tau } (\alpha
\psi )
=
\alpha ^{\dagger}\alpha \psi ^{\dagger} \partial _{\tau }\psi   +
\alpha ^{\dagger}(\partial _{\tau }\alpha ) \psi ^{\dagger} \psi
\end{eqnarray}
With our consideration of `complete chains of Grassmann numbers'
we can write the partition function:
\begin{eqnarray}
Z&=& \int \left( \prod _{n=1} ^{N } d \alpha ^{\dagger} _{n}
d \alpha _{n}  d \psi ^{\dagger}_{n} d \psi_{n} \right)
\left(
\prod _{n=1} ^{N }\alpha ^{\dagger} _{n} \alpha  _{n} \psi _{n}
\psi ^{\dagger} _{n}
-\prod _{n=1} ^{N }\alpha ^{\dagger} _{n} \alpha  _{n-1}
\psi ^{\dagger} _{n}\psi _{n-1}\left( 1+\mu \epsilon \right)
\right) \nonumber
\end{eqnarray}
We note that
each of the two chains possesses a complete product
of the complex fields $\alpha _{n}$.
Therefore we can integrate out the  $\alpha _{n}$
and we get the known  result
\begin{eqnarray}
Z&=& const \int \left( \prod _{n=1} ^{N }  d \psi ^{\dagger}_{n} d
\psi_{n}\right)
\left(
\prod _{n=1} ^{N }\psi _{n}
\psi ^{\dagger} _{n}
-\prod _{n=1} ^{N }
\psi ^{\dagger} _{n}\psi _{n-1}\left( 1+\mu \epsilon \right)
\right)
\end{eqnarray}
which can be re-exponentiated:
\begin{eqnarray}
Z= \int {\cal D }\left[ \psi ^{\dagger}( \tau ),\psi (\tau ) \right]
\exp \left[ -\int _{0} ^{\beta } d \tau \ ( \psi ^{\dagger} \partial _{\tau
}\psi -
\mu \psi ^{\dagger} \psi ) \right]
\end{eqnarray}
In this way we have shown   the additional field  to be superfluous.

\subsection{Particle Hole Transformation}
\label{Anhang5}

The particle hole transformation is a special case of
an unitary transformation for electronic states.
On the operator level it reads:
\begin{eqnarray}
C \rightarrow B^{\dagger} \,   , \ \ C ^{\dagger}\rightarrow B
\end{eqnarray}
Under this transformation the trivial
Hamilton operator $H =- \mu   C^{\dagger} C $ transforms to
$H\rightarrow H' = -\mu B B ^{\dagger} = -\mu \1 +\mu
B^{\dagger} B$.
So, why not introduce  the analogous unitary transformation
in the path integral formalism
\begin{eqnarray}
\psi _{n}  \rightarrow \psi^{\dagger}_{n} \,   , \ \  \psi
^{\dagger}_{n}\rightarrow
\psi _{n}\ \ \ \  ?
\end{eqnarray}

In the path integral $\mu \psi ^{\dagger} \psi $ transforms into $-\mu \psi
^{\dagger}
\psi $.
However, we expect the particle hole transformed result to be $\mu -\mu
\psi ^{\dagger} \psi $.
Obviously, we have to do further considerations to find
a Lagrangian analogous to a particle hole transformed Hamiltonian.  \\

The path integral consists of  two complete Grassmann chains.
One of them is the particle chain $\prod \psi ^{\dagger}_{l}\psi _{l-1}$ and
the other is the
hole chain $\prod \psi _{l}\psi  ^{\dagger}_{l }$.
The particle hole transformation should exchange the two  chains.
Therefore the particle hole transformation is not unitary, but `almost
unitary':
\begin{eqnarray}
\psi^{\dagger}_{n} \rightarrow \psi _{n} \left( 1- \mu \epsilon \right) \, ,
\ \ \
\psi _{n-1} \rightarrow \psi^{\dagger}_{n}\label{TLTrafo}
\end{eqnarray}
and the Jacobian is\footnote{The Jacobian for the Grassmann numbers is defined
inversely in comparison
to the complex numbers.}:
\begin{eqnarray}
|J|  = \prod _{n=1} ^{N } \left( 1+\mu \epsilon \right)
=e^{\beta \mu }
\end{eqnarray}
The Berry phase  transforms to (first order in $\epsilon $):
\begin{eqnarray}
&&\lim _{N\rightarrow \infty} \sum _{n=1}^{N}
\psi _{n}^{\dagger} (\psi _{n}-\psi _{n-1})
\nonumber \\
&& \rightarrow
\lim _{N\rightarrow \infty} \sum _{n=1}^{N}
\psi _{n} (1-\mu \epsilon )(\psi _{n+1} ^{\dagger}-\psi _{n}^{\dagger})
=\lim _{N\rightarrow \infty} \sum _{n=1}^{N}
\psi _{n}^{\dagger}(\psi _{n} -\psi _{n-1}) +{\cal O }(\epsilon ^{2})
\end{eqnarray}
and the partition function accordingly:
\begin{eqnarray}
Z&=&\int \left(\prod _{n=1} ^{N }  d \psi ^{\dagger}_{n} d \psi_{n} \right)
\exp \left[ -\sum _{n=1} ^{N }  \epsilon
\left( \psi ^{\dagger} _{n} \left( \psi _{n } - \psi _{n-1} \right) -
\mu \psi ^{\dagger} _{n}  \psi _{n-1}
\right)  \right]
\nonumber \\
&\rightarrow & |J|\int \left( \prod _{n=1} ^{N }
d \psi ^{\dagger}_{n} d \psi_{n} \right)
\exp \left[ -\sum _{n=1} ^{N }  \epsilon
\left(
\psi ^{\dagger} _{n} \left( \psi _{n } - \psi _{n-1} \right)
-\mu \psi _{n} \psi _{n}^{\dagger} +{\cal O} (\epsilon )
\right) \right]
\end{eqnarray}
If we take the continuum limit, i.e.\ keep the first order in $\epsilon$ only,
we find:
\begin{eqnarray}
Z&=& e^{\mu \beta }\int  \left( \prod _{n=1} ^{N }d \psi ^{\dagger}_{n} d
\psi_{n} \right)
\exp \left[ -\sum _{n=1} ^{N }  \epsilon
\left(
\psi _{n} ^{\dagger} ( \psi _{n} - \psi _{n-1})
+\mu \psi _{n}^{\dagger} \psi _{n}
\right) \right]
\label{aabbc}
\end{eqnarray}
So, transformation (\ref{TLTrafo}) produced the expected
$e^{\mu \beta }$-prefactor.

\section{Resolution of Unity for Infinite Interaction?}
\label{Widerspruchsbeweis}

In this appendix we will investigate the possibility to
resolve the unity operator with normalized coherent
states which exist in the constrained Fock space.
We remind the reader that the resolution of the unity operators
with such coherent states would
result  in a well-defined Berry phase in the continuum limit.
For this purpose we will explicitly construct coherent states
with either one or two complex Grassmann fields which are
clearly normalized.

However, it will be proved by contradiction that
neither of them will resolve the unity operator (further discussions in
\cite{et}) -- assuming a sufficiently `simple' measure which is
not a sum of several
Grassmann valued terms.

\subsection{Coherent state with two complex Grassmann numbers}

We write the  normalized coherent state in exponential form as in
Perelomov \cite{Perelomov} (cf. (\ref{DefkohZust})).
The bosonic  operators will  acquire a complex prefactor, and the
fermionic   a Grassmann prefactor.
In the following formula $\alpha $ is a complex field, but
$\beta $ is proportional to a Grassmann number $\psi _{\+}$
and $\gamma $ proportional to a Grassmann number $\psi _{\-}$.
The reference state  is chosen to be be $|0\rangle $.
\begin{eqnarray}
|g\rangle &=& \exp \left[
\alpha X^{\+ \- } + \beta X^{\+ 0} +\gamma X^{\- 0}
-\alpha ^{*}  X^{\- \+ } + \beta ^{*} X^{0\+ } +\gamma ^{*} X^{0\- }
\right]  | 0 \rangle \nonumber \\
&=&
|0 \rangle
+ \beta |\+  \rangle +\gamma  |\-  \rangle \nonumber \\
&+&
\frac{1}{2!} \left\{
\alpha \gamma |\+  \rangle - \alpha ^{*} \beta |\-  \rangle
-(\beta ^{*}\beta + \gamma ^{*} \gamma )|0 \rangle
\right\}
\nonumber \\
&+&
\frac{1}{3!} \left\{
(-\alpha \alpha ^{*} \beta -\beta  \gamma ^{*} \gamma )|\+  \rangle
+(\alpha ^{*} \alpha \gamma - \gamma \beta ^{*}\beta )|\- \rangle
+(\gamma ^{*} \alpha ^{*} \beta -\beta ^{*}\alpha \gamma )|0\rangle
 \right\} \nonumber \\
&+& \ldots
\end{eqnarray}
The construction scheme is found  by expanding to eighth order.
The coherent state then is:
\begin{eqnarray}
& & \left\{ \beta \frac{\sin (|\alpha |)}{ |\alpha | }
+ \gamma \frac{1-\cos ( |\alpha |)}{ \alpha ^{*} }
+\beta \gamma ^{*} \gamma \frac{\sin (|\alpha |)-|\alpha |}{ |\alpha | ^{3}}
\right\} |\+  \rangle \nonumber \\ & &
+
\left\{ \gamma \frac{\sin (|\alpha |)}{ |\alpha | }
- \beta  \frac{1-\cos ( |\alpha |)}{ \alpha  }
+ \gamma \beta  ^{*} \beta \frac{\sin (|\alpha |)-|\alpha |}{ |\alpha | ^{3}}
\right\} |\-  \rangle \nonumber \\ & &
+\left\{
1-\gamma ^{*} \beta \frac{\sin (|\alpha |)-|\alpha |}{ \alpha |\alpha |}
+ \beta  ^{*} \gamma \frac{\sin (|\alpha |)-|\alpha |}{ |\alpha
|\alpha  ^{*} } \right.  \nonumber \\ & & \left.
-\left( \beta  ^{*} \beta + \gamma  ^{*}\gamma \right)
\frac{1-\cos ( |\alpha |)}{ |\alpha |^{2} }
-2 \beta  ^{*} \beta \gamma  ^{*}\gamma
\frac{1-\cos ( |\alpha |) - |\alpha | ^{2} /2 }{ \alpha  }
\right\} |0 \rangle
\end{eqnarray}
which is proved by induction.
This state is normalized.
Next the  resolution of the unity operator is tackled.
Therefore we have to take the projector onto the coherent state and
integrate with some   unknown measure.
This measure is either a constant, a single product of Grassmann numbers,
or a sum of such terms. The single product is ruled out, since it destroys
prefactors of the diagonal Hubbard operators in the projection operator.
The sum of products yields the measure of section \ref{erer}
as the only possible choice.
Finally, the case that the measure is a constant is investigated below.

The $ \arg (\alpha ) $ phase integration over $  \alpha   $
is supposed to
yield  0 in order to destroy the terms with
off-diagonal Hubbard operators. The
remaining terms, with diagonal Hubbard operators,  contain  integrals
as prefactors which
should equal 1:
\begin{eqnarray}
X^{\+ \+ } &:& \! \!
\int 2 \beta \beta  ^{*}\gamma \gamma  ^{*}\frac{\sin (|\alpha |)}{ |\alpha | }
\frac{|\alpha |-\sin (|\alpha |)}{ |\alpha | ^{3}} \stackrel{!}{=}1 \label{e1}
\\
X^{\- \- } &:& \! \!
\int 2 \beta \beta  ^{*}\gamma \gamma  ^{*}\frac{\sin (|\alpha |)}{ |\alpha | }
\frac{|\alpha |-\sin (|\alpha |)}{ |\alpha | ^{3}}\stackrel{!}{=}1 \label{e2}\\
X^{0 0  } &:& \! \!
\int  \beta \beta  ^{*}\gamma \gamma ^{*}
\left[
4\frac{\cos (|\alpha |)-1+|\alpha | ^{2} /2 }{|\alpha |^{4} }
-2\frac{\left( |\alpha |  -\sin  (|\alpha |)      \right)
^{2}}{|\alpha | ^{4}}
+ 2\frac{\left( \cos  (|\alpha |)  -1    \right)
^{2}}{|\alpha | ^{4}}
\right] \stackrel{!}{=} 1 \label{e3}
\end{eqnarray}
The square brackets can be written as:
\begin{eqnarray}
\left[  ....
\right] = \frac{
4\sin (|\alpha |) (|\alpha | - \sin (|\alpha |) )}{ |\alpha | ^{4}  }
\end{eqnarray}
Now,   adding  the first two equations (\ref{e1}),(\ref{e2}) results in
a contradiction to    equation (\ref{e3}).

To complete the proof one can choose another reference state, for example
$|\+ \rangle $ or $|\- \rangle $.
Carrying through the same steps as before one obtains    a  similar
contradiction.

\subsection{Coherent state with one  complex Grassmann number}
\label{klkl}

Analogously, we try to find a resolution of unity with one Grassmann number.
For this purpose we take the calculation from the previous section and
substitute  $\psi _{1} =\psi _{2}$. Then   the coherent state
is calculated to be:
\begin{eqnarray}
& & \left\{ \beta \frac{\sin (|\alpha |)}{ |\alpha | }
+ \gamma \frac{1-\cos ( |\alpha |)}{ \alpha ^{*} }
\right\} |\+  \rangle \nonumber \\ & &
+
\left\{ \gamma \frac{\sin (|\alpha |)}{ |\alpha | }
- \beta  \frac{1-\cos ( |\alpha |)}{ \alpha  }
\right\} |\-  \rangle   \\ & &
+\left\{
1-\gamma ^{*} \beta \frac{\sin (|\alpha |)-|\alpha |}{ \alpha |\alpha |}
+ \beta  ^{*} \gamma \frac{\sin (|\alpha |)-|\alpha |}{ |\alpha
|\alpha  ^{*} }
-\left( \beta  ^{*} \beta + \gamma  ^{*}\gamma \right)
\frac{1-\cos ( |\alpha |)}{ |\alpha |^{2} }
\right\} |0 \rangle \nonumber
\end{eqnarray}
Again, this state is normalized.
To resolve the unity we have to integrate with an unknown measure.
Using the same procedure as before we get:
\begin{eqnarray}
X^{\+ \+ } & :  &
\int \beta \beta  ^{*}\frac{\sin ^{2}(|\alpha |)}{ |\alpha |^{2} }
+\gamma \gamma  ^{*}\frac{(1-\cos(|\alpha |) ) ^{2}}{ |\alpha |^{2} }
\stackrel{!}{=}1 \\
X^{\- \- } & :  &
\int  \gamma \gamma  ^{*}\frac{\sin ^{2}(|\alpha |)}{ |\alpha |^{2} }
+ \beta \beta ^{*}\frac{(1-\cos(|\alpha |) ) ^{2}}{ |\alpha |^{2} }
\stackrel{!}{=}1 \\
X^{0 0 } & :  &
\int 2( \beta \beta  ^{*}+\gamma \gamma ^{*} )
\frac{(\cos(|\alpha |) -1) ^{2}}{ |\alpha |^{2} }
\stackrel{!}{=} 1 \nonumber \\
\end{eqnarray}
Adding the first two lines results in  a contradiction to the last.

\section{Slave Particle Constraints}

To contrast the projection  methods discussed in this paper with
slave boson techniques,  we sketch the treatment of  slave boson  constraints
within a path integral formulation \cite{RN,Read2}.
The trace in the partition function
is  constrained to some subspace characterized by $q=0$,
\begin{eqnarray}
Z= \Tr \left[  e^{-\beta H} |q=0\rangle \langle q=0|\right]
\end{eqnarray}
where   $q$
labels the eigenvalues of    operator $Q$.
We may write
\begin{eqnarray}
Z= \Tr \left[  e^{-\beta H} \delta (Q)\right]
=\int \frac{d \lambda ' }{ 2 \pi}\Tr
\left[ e^{-\beta H} e^{i\lambda ' Q} \right]
\end{eqnarray}
The delta function makes sense if we evaluate the trace
in the basis of eigenstates of Q. We now assume that Q commutes with $H$
to get\footnote{e.g.\ a path integral for $H=\sum _{<i,j>} \sum _{\sigma }
t_{ij} C^{\dagger }_{i\sigma }C_{j\sigma }$
with constraint $\lambda ' Q= \lambda ' (\sum _{i} n _{i\+}n_{i\downarrow}-0)$
is invalid because $Q$ does not commute with $H$.}
\begin{eqnarray}
Z= \int _{-\pi }^{\pi }\frac{d \lambda '}{ 2 \pi}\Tr
\left[e^{-\beta H +i\lambda ' Q }\right]
\end{eqnarray}
The integral runs from $-\pi$ to $\pi$ because the eigenvalues
of the number operator $Q$ are supposed to be  integers.
Now, the coherent state functional integral may be set up in the usual way
($\lambda =\lambda '/\beta $),
\begin{eqnarray}
Z&=& \int _{-\pi T}^{\pi T}\frac{\beta d \lambda }{ 2 \pi}\Tr
\left[e^{-\beta H (\lambda) }\right] \nonumber \\ &=&
\lim _{n\rightarrow \infty }
\int _{-\pi T}^{\pi T}\frac{\beta d \lambda }{ 2 \pi}\Tr
\left[ \1 :e^{-\Delta \tau H(\lambda )} :\1: e^{-\Delta \tau H(\lambda )}:\1:
e^{-\Delta \tau H(\lambda )} :  ...  \1   \right]
\end{eqnarray}
where $H(\lambda )= H-i\lambda Q $.
We may expand the exponentials in this Trotter formula only because
$\beta \lambda $ is limited to the interval $\left[-\pi, \pi \right]$.
It remains to express the unities by a closure relation for coherent states.

We should  draw the readers attention to the following two facts concerning
the constraint term:

(i) The constraint is not implemented  at each time step separately
-- nor at the initial time step.
It was implemented at the operator level and is now `smeared' over all time
steps.
This contrasts with  the projection methods introduced in the main body of the
paper.

(ii) $\lambda $ is a parameter (one for each site), and not a (time-dependent)
field.
It is not correct to make $\lambda $ time dependent $(\lambda \rightarrow
\lambda _{n})$. However, since $Q$ is a    number operator, say
$i\lambda Q= i \lambda  \left( \sum _{\alpha }  a ^{\dagger }_{\alpha }
a_{\alpha } -1 \right)\rightarrow i \lambda \sum _{\alpha }
\Phi ^{*} _{\alpha _{n}}\Phi _{\alpha _{n-1}} - i \lambda  $,
we may perform a gauge transformation of the fields
$\Phi _{\alpha _{n}}\rightarrow e^{i \chi _{n}}\Phi _{\alpha _{n}}$
to generate additional terms in the Berry  phase:
\begin{eqnarray}
\Phi ^{*} _{\alpha _{n}}\left( \Phi _{\alpha _{n}} - \Phi _{\alpha _{n-1}}
\right) &\rightarrow&
\Phi ^{*} _{\alpha _{n}} e^{-i \chi _{n}} \left(
\Phi _{\alpha _{n}}e^{i \chi _{n}} - \Phi _{\alpha _{n-1}}
e^{i \chi _{n-1}}\right) \nonumber \\
&=&\Phi ^{*} _{\alpha _{n}}\left( \Phi _{\alpha _{n}} - \Phi _{\alpha _{n-1}}
\right)
+\Phi ^{*} _{\alpha _{n}}\Phi   _{\alpha _{n-1}} \left(
1- e^{i( \chi _{n-1}-\chi _{n} )} \right)
\end{eqnarray}
The additional term  may be written  as $i \Phi ^{*} _{\alpha  }\Phi  _{\alpha
}
\partial _{\tau } \chi $,
if $\partial _{\tau } \chi $ stays finite (i.e.\ if $  \chi _{n} -\chi _{n-1}
={\cal O} (\Delta \tau )$).
This gauge field has to obey the boundary condition $\chi (0) - \chi (\beta )
=2\pi m $
to keep $\exp (iS)$ invariant, and
$\lambda \rightarrow \lambda +i\dot{\chi}(\tau) =: \lambda (\tau )$
(see the discussion in   \cite{Fradkin}).
Now we sum all paths of $\lambda (\tau )$ in order to obtain a functional
integration of the Lagrange multiplier field $\lambda (\tau )$.
Interpretation of this procedure is as follows:
The static component of the field $\lambda (\tau )$ enforces the constraint
whereas the Fourier components may be gauged away.

\end{document}